\pdfoutput=1

\documentclass[11pt,twoside,a4paper,cmspaper,final,collab]{cms-tdr}
\def\svnVersion{486625P}\def\cmsCernNoTag{CERN-EP-2018-182}\def\cmsCernDate{\today}\def\cmsMessage{Published in the Journal of High Energy Physics as \href{http://dx.doi.org/10.1007/JHEP01(2019)051}{\doi{10.1007/JHEP01(2019)051.}}}

\begin{document}\cmsNoteHeader{B2G-17-006}

\hyphenation{had-ron-i-za-tion}
\hyphenation{cal-or-i-me-ter}
\hyphenation{de-vices}
\RCS$HeadURL: svn+ssh://svn.cern.ch/reps/tdr2/papers/B2G-17-006/trunk/B2G-17-006.tex $
\RCS$Id: B2G-17-006.tex 407633 2017-05-31 13:21:24Z cgalloni $

\providecommand{\cmsLeft}{left\xspace}
\providecommand{\cmsRight}{right\xspace}
\newcommand{\PV}{\ensuremath{\mathrm{V}}\xspace}
\newlength\cmsTabSkip\setlength{\cmsTabSkip}{1ex}
\newcommand{\Vjets}{\ensuremath{\PV\text{+jets}}\xspace}
\newcommand{\PVpr}{\ensuremath{\mathrm{V}'}\xspace}

\cmsNoteHeader{B2G-17-006}

\title{Search for heavy resonances decaying into two Higgs bosons or into a Higgs boson and a {\PW} or \PZ boson in proton-proton collisions at 13\TeV}

\date{\today}

\abstract{A search is presented for massive narrow resonances decaying either into two Higgs bosons, or into a Higgs boson and a {\PW} or \PZ boson. The decay channels considered are $\PH\PH \to {\bbbar}\tau^{+}\tau^{-}$ and $\PV\PH \to {\qqbar}\tau^{+}\tau^{-}$, where \PH denotes the Higgs boson, and \PV denotes the {\PW} or \PZ boson. This analysis is based on a data sample of proton-proton collisions collected at a center-of-mass energy of 13\TeV by the CMS Collaboration, corresponding to an integrated luminosity of 35.9\fbinv. For the \TeV-scale mass resonances considered, substructure techniques provide ways to differentiate among the hadronization products from vector boson decays to quarks, Higgs boson decays to bottom quarks, and quark- or gluon-induced jets.
Reconstruction techniques are used that have been specifically optimized to select events in which the tau lepton pair is highly boosted. The observed data are consistent with standard model expectations and upper limits are set at 95\% confidence level on the product of cross section and branching fraction for resonance masses between 0.9 and 4.0\TeV. Exclusion limits are set in the context of bulk radion and graviton models: spin-0 radion resonances are excluded below a mass of 2.7\TeV at 95\% confidence level.
In the spin-1 heavy vector triplet framework, mass-degenerate \PWpr and \PZpr resonances
with dominant couplings to the standard model gauge bosons are excluded below a mass of 2.8\TeV at 95\% confidence level.
These are the first limits for massive resonances at the \TeV scale with these decay channels at $\sqrt{s}=13\TeV$.
}

\hypersetup{
pdfauthor={CMS Collaboration},
pdftitle={Search for heavy resonances decaying into two Higgs bosons or
 into a Higgs boson and a W or Z boson in proton-proton collisions at 13 TeV},
pdfsubject={CMS},
pdfkeywords={CMS, physics, Higgs physics, beyond standard model}}

\maketitle

\section{Introduction}
\label{sec:introduction}

\begin{figure}[!htb]
  \centering
  \includegraphics[width=.32\textwidth]{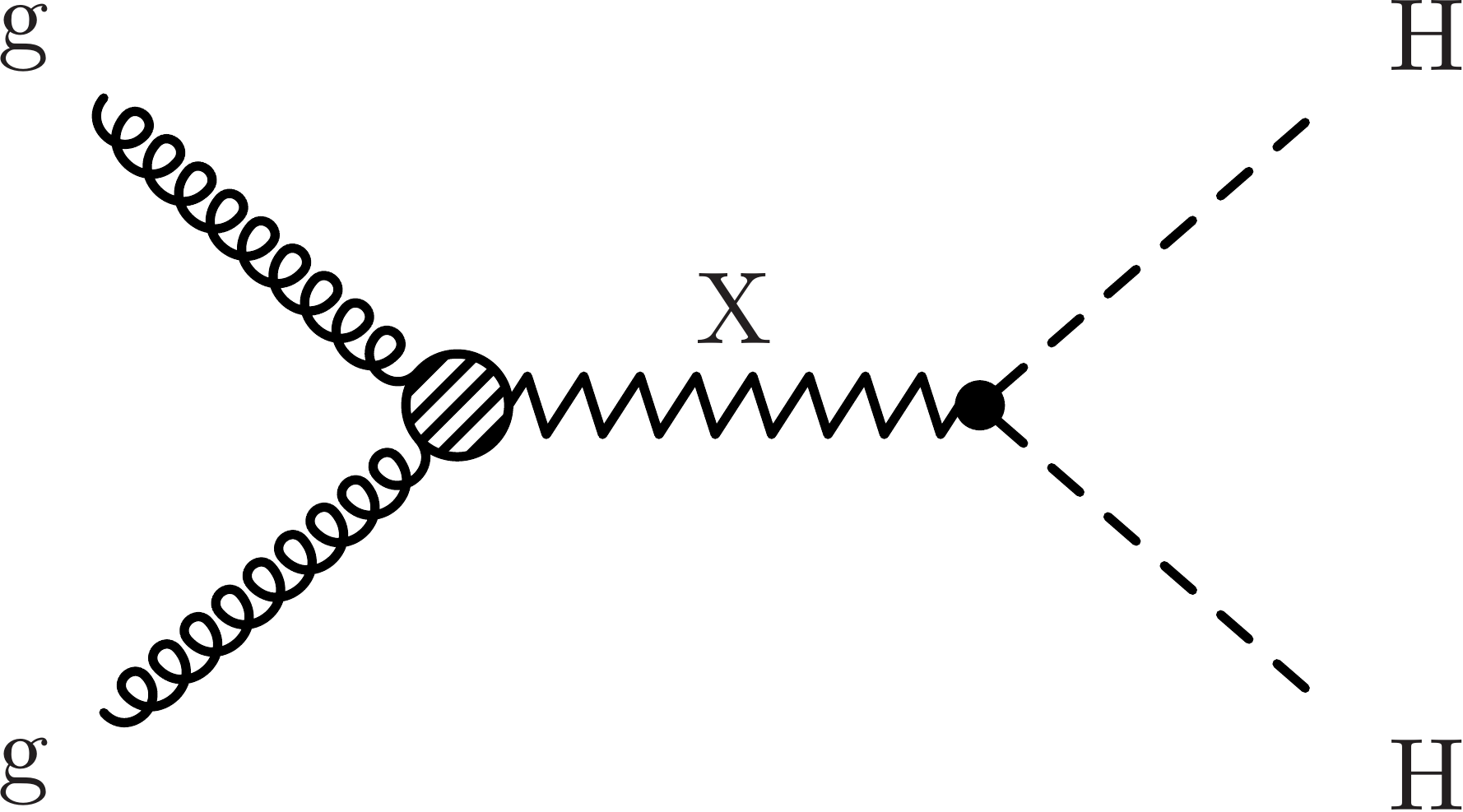} \hspace{1cm}
  \includegraphics[width=.32\textwidth]{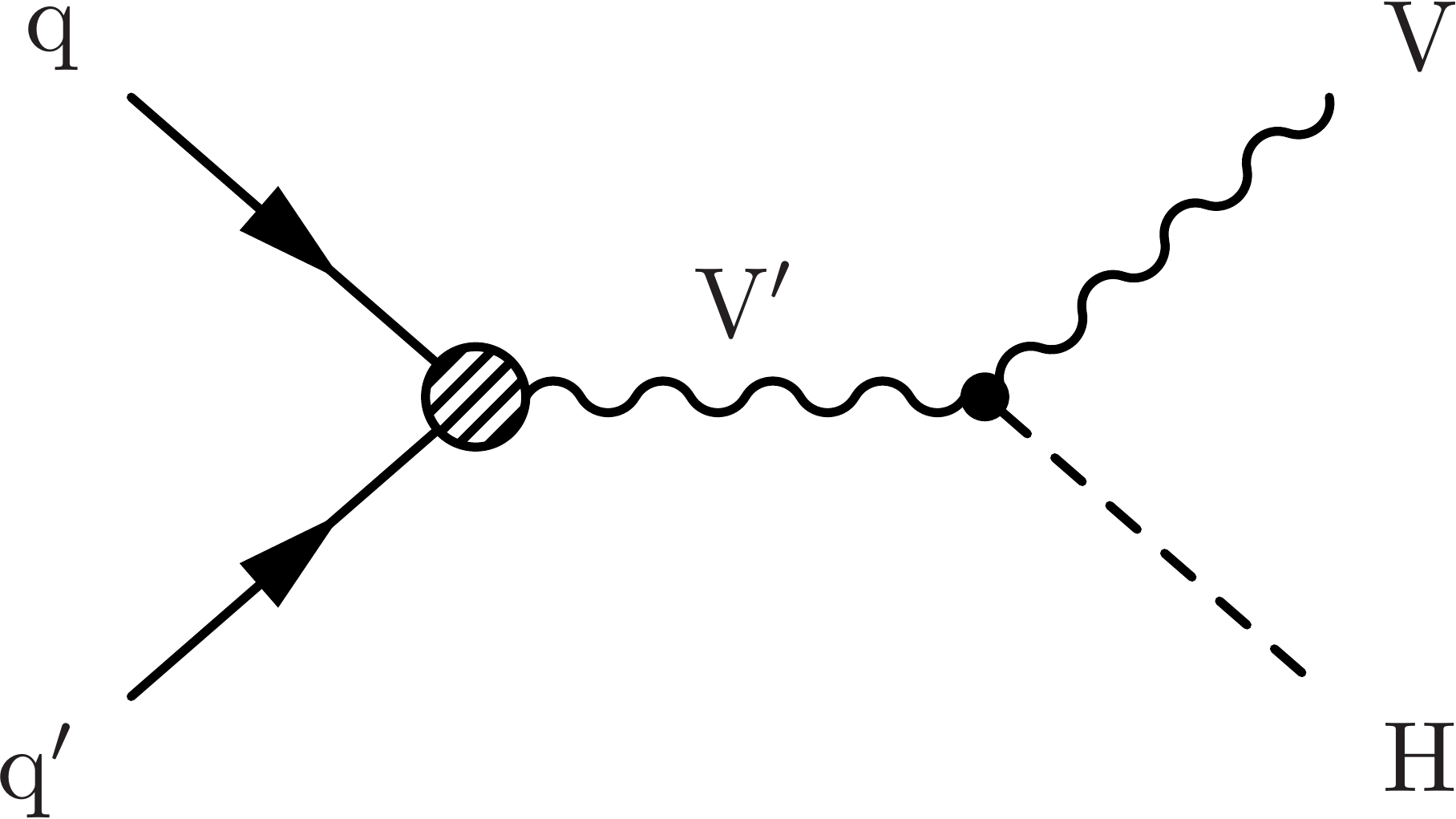}

  \caption{Feynman diagrams for the production of a spin-0 radion or a spin-2 graviton X that decays to two Higgs bosons (\cmsLeft), and the production of a heavy vector boson \PVpr (\PWpr or \PZpr) that decays to a vector boson and a Higgs boson (\cmsRight).}
  \label{fig:Feynman_diagram}
\end{figure}

Heavy resonances that decay to $\PH\PH$, $\PV\PV$, or $\PV\PH$, where \PH denotes the Higgs boson, and \PV denotes a {\PW} or \PZ boson, are motivated by theories beyond the standard model (SM) that address the large difference between the electroweak and gravitational scales. These heavy particles arise as Kaluza--Klein (KK) excitations of spin-0 radions~\cite{Goldberger:1999uk,DeWolfe:1999cp,Csaki:1999mp}, and as spin-2 gravitons predicted in models based on Randall--Sundrum warped extra dimensions~\cite{Randall:1999ee,Randall:1999vf}, with the gravitons propagating in the entire five-dimensional bulk~\cite{Agashe:2007zd, Fitzpatrick:2007qr, Antipin:2007pi}. Heavy spin-1 \PWpr and \PZpr particles that decay to $\PV\PV$ and $\PV\PH$ are also postulated in composite Higgs models~\cite{Bellazzini:2014yua,CHM2,Composite2,Greco:2014aza}, little Higgs models~\cite{Schmaltz:2005ky,ArkaniHamed:2002qy}, and in the sequential SM (SSM)~\cite{Altarelli}. The models containing new spin-1 states are generalized in the heavy vector triplet (HVT) framework~\cite{Pappadopulo:2014qza}. All of the new hypothetical particles with spins of 0, 1, or 2 can be produced at the CERN LHC, via the processes depicted in the Feynman diagrams of Fig.~\ref{fig:Feynman_diagram}.

The bulk graviton model has two free parameters: the mass of the first KK excitation of the spin-2 boson, denoted as the KK bulk graviton, and the ratio $\tilde{k} \equiv k/\overline{M}_{\mathrm{Pl}}$, where $k$ is the unknown curvature scale of the extra dimension and $\overline{M}_{\mathrm{ Pl}} \equiv M_{\mathrm{Pl}}/\sqrt{8\pi}$ is the reduced Planck mass. Searches for radions in this model can be described in terms of the radion mass and the ultraviolet cutoff of the theory $\Lambda_{\mathrm{R}}$ \cite{Gouzevitch:2013qca}. The HVT model is formulated in terms of four parameters: the mass of the new vector bosons, their coupling coefficient to fermions $c_{\mathrm{F}}$, their coupling coefficient to the Higgs boson and longitudinally-polarized SM vector bosons $c_{\PH}$, and the strength of the new vector boson interaction $g_{\PV}$.
In the HVT framework two scenarios are considered and henceforth referred to as model A and model B, depending on the couplings of the new physics resonance to the SM particles. In model A ($g_{\PV}=1$, $c_{\PH}=-0.556$, $c_{\mathrm{F}}=-1.316$), the coupling strengths to the SM bosons and fermions are comparable and the new particles decay primarily to fermions. In model B ($g_{\PV}=3$, $c_{\PH}=-0.976$, $c_{\mathrm{F}}=1.024$), the coupling to the SM fermions are small, and the branching fraction of the new resonance to SM bosons is nearly 100\% \cite{Pappadopulo:2014qza}.
Searches for diboson resonances have previously been performed in several final states, placing lower limits on the masses of such resonances above the \TeV scale \cite{Aaboud:2017cxo, Sirunyan:2018qob, Aaboud:2017ahz,Sirunyan:2018ivv, Sirunyan:2018hsl, PhysRevD.97.072006, Aaboud:2017itg, Aaboud:2017eta, Aaboud:2017fgj, Aaboud:2017gsl, Aaboud:2018ohp, Sirunyan:2017jtu, Sirunyan:2017isc,  Aaboud:2018knk,Sirunyan:2017nrt, Aaboud:2018bun}.

This paper presents a search for resonances with masses above 900\GeV decaying into \PH{}\PH{} or \PV{}\PH{}.
The final states analyzed are $\PH\PH \to \bbbar \TT$ and $\PV\PH \to \qqbar\TT$. Events are classified as ``semileptonic'', denoted as $\ell\tauh$, if one {\Pgt} lepton decays into a lighter lepton ($\ell$), denoting either a muon or an electron, and two neutrinos, and the other decays hadronically ($\tauh$) into hadrons and a neutrino. Events are categorized as ``fully-hadronic'', denoted as $\tauh\tauh$, if both \Pgt{} leptons decay hadronically.
The analysis aims to reconstruct the diboson decay products in order to search for a narrow local enhancement in the diboson invariant mass spectrum with a width insignificant with respect to the experimental resolution. This search complements and significantly extends the reach of the CMS search for $\PZ\PH\to \qqbar\TT$ resonances with data collected at $\sqrt{s}=8\TeV$ \cite{Khachatryan:2015ywa}, by using a larger data sample collected at $\sqrt{s}=13\TeV$, a more complex categorization of the final states, and an increased number of signal models.
Resonances of masses below the \TeV scale have been excluded by other searches in similar final states \cite{Sirunyan:2017djm, Aaboud:2018sfw}.

Since the resonances under study have masses of the order of a few \TeV, the bosons resulting from their decays have transverse momenta (\pt) of at least several hundred GeV. As a consequence, the final decay products are collimated such that the hadronically decaying bosons cannot be resolved using standard jet algorithms. Dedicated techniques, called \PV tagging and \PH tagging, are applied to exploit the substructure of these large-cone jets to resolve the hadronically decaying vector and Higgs bosons.

For the Higgs boson decaying to a pair of \Pgt{} leptons, the decay products are also close in angular separation. The \Pgt{} lepton reconstruction and identification techniques described in Ref.~\cite{Khachatryan:2015ywa} are therefore adopted to achieve optimal signal significance for this particular event topology.

\section{The CMS detector}
\label{sec:detector}

The central feature of the CMS apparatus is a superconducting solenoid of 6\unit{m} internal diameter, providing a magnetic field of 3.8\unit{T}. Within the solenoid volume are a silicon pixel and strip tracker, a lead tungstate crystal electromagnetic calorimeter (ECAL), and a brass and scintillator hadron calorimeter (HCAL), each composed of a barrel and two endcap sections. Forward calorimeters extend the pseudorapidity ($\eta$) coverage provided by the barrel and endcap detectors. Muons are measured in gas-ionization detectors embedded in the steel flux-return yoke outside the solenoid. The CMS two-level trigger system~\cite{Khachatryan:2016bia} reduces the event rate from the bunch crossing rate of 40\unit{MHz} down to around 1\unit{kHz} for data storage.

A more detailed description of the CMS detector, together with a definition of the coordinate system used and the kinematic variables, can be found in Ref.~\cite{Chatrchyan:2008aa}.

\section{Data sample and simulation}
\label{sec:samples}

The data sample analyzed in this search corresponds to an integrated luminosity of 35.9\fbinv, collected in 13\TeV proton-proton collisions with the CMS detector during the 2016 data taking period. The signal processes $\Pp\Pp \to\mathrm{X}\to \PV\PH \to \qqbar\TT$ and $\Pp\Pp \to\mathrm{X}\to \PH\PH \to \bbbar \TT$ are simulated at leading order (LO) using the \MGvATNLO v2.2.2~\cite{Alwall:2014hca} Monte Carlo (MC) event generator, for resonance masses between 900 and 4000\GeV, where the Higgs boson is forced to decay to $\tau$ pairs and the other boson to a pair of quarks. The signal processes where $\Pp\Pp \to\mathrm{X}\to \PH\PH \to \bbbar \PV\PV$ and $\Pp\Pp \to\mathrm{X}\to \PV\PH \to \qqbar \PV\PV$ are also considered, in which $\PV\PV \to 2\ell2\nu$ or $2\tau2\nu$, as they can yield final states similar to those of the primary signal process. The natural width of the resonance is assumed to be smaller than the experimental resolution of its reconstructed mass, as consistent with the benchmark radion, graviton and HVT models.

The SM background processes are generated using MC simulation.
The $\PZ/\gamma^{*}$+jets events and the {\PW}+jets events are simulated at LO with the \MGvATNLO generator.
The \POWHEG v2 generator is used to simulate \ttbar and single top quark production at next-to-leading order~\cite{Nason:2004rx,Frixione:2007vw,Alioli:2010xd,Alioli:2011as}. The LO \PYTHIA 8.205~\cite{Sjostrand:2014zea} generator is used for SM diboson ($\PW\PW$, $\PW\PZ$, or $\PZ\PZ$) and multijet events. For all signal and background samples, showering and hadronization are modeled using \PYTHIA, \Pgt{} lepton decays are described using \textsc{tauola} 1.1.5~\cite{Davidson:2010rw}, and the response of the detector is simulated using \GEANTfour~\cite{Agostinelli:2002hh}. Additional collisions in the same or adjacent bunch crossings (pileup) are superimposed onto the hard scattering processes, with the pileup vertex multiplicity distribution adjusted to match that of data.

\section{Event reconstruction}
\label{sec:eventReconstruction}

The particle flow (PF) event algorithm~\cite{Sirunyan:2017ulk} reconstructs and identifies each individual particle through an optimized combination of information from the various elements of the CMS detector.
The energy of each electron is determined from the electron momentum as determined by the tracker, the energy of the corresponding ECAL cluster, and the energy sum of all bremsstrahlung photons spatially compatible with originating from the electron track. The energies of muons are obtained from the curvature of the corresponding tracks.
The energies of charged hadrons are determined from a combination of their momentum measured in the tracker and the matching ECAL and HCAL energy deposits, corrected for zero-suppression effects and for the response function of the calorimeters to hadronic showers. Finally, the energies of neutral hadrons are obtained from the corresponding corrected ECAL and HCAL energies.

The identified particles are clustered into jets using the anti-\kt algorithm~\cite{Cacciari:2008gp}, implemented in the {\FASTJET} package~\cite{Cacciari:2011ma}. Two distance parameters are used in the analysis, 0.4 and 0.8, yielding jet collections referred to as AK4 and AK8 jets, respectively. The AK4 jets are used primarily to reject or select events with top quarks, while the larger AK8 jets are used to identify and contain hadronically decaying {\PW}, \PZ, and Higgs boson candidates. The charged-hadron subtraction (CHS) algorithm for mitigating pileup~\cite{Sirunyan:2017ulk} discards charged particles not originating from the primary vertex (PV). The PV is defined as the vertex with the largest $\pt^{2}$ sum of the physics objects. Here, the physics objects are the AK4 jets,
and the associated missing transverse momentum, which is taken as the negative vector sum of the \pt of the jets.
The residual contamination from neutral pileup particles is estimated to be proportional to the event energy density and the jet area, and is removed from the jet energy calculation.
The jet momentum is determined as the vectorial sum of all particle momenta in the jet, and is found from simulation to be within 5 to 10\% of the true momentum over the entire \pt spectrum and detector acceptance.
Jet energy corrections are obtained from simulation, and are confirmed with in situ measurements of the energy balance in dijet, multijet, $\gamma$+jets, and leptonically decaying {\PZ}+jets events~\cite{Khachatryan:2016kdb}. Additional selection criteria are applied to each event to remove spurious jet-like features originating from isolated noise patterns in certain HCAL regions~\cite{CMS:2016ljj}. The AK4 and AK8 jets must have $\pt>20$ and $\pt>200\GeV$, respectively, and $\abs{\eta}<2.4$ to be considered in the subsequent steps of the analysis.

To determine the jet mass and the substructure variables used in the identification of the hadronic decays of bosons, the so-called pileup per particle identification (PUPPI) algorithm~\cite{Bertolini2014} is applied to AK8 jets, instead of the CHS algorithm, to retain better stability of the substructure variables in events with a large amount of pileup. The PUPPI algorithm uses the local distribution of particles, event-pileup properties, and tracking information to compute a weight describing the likelihood for each particle to originate from a pileup interaction. The weights are used to rescale the particle four-momenta, superseding the need for further jet-based corrections.
These particles are subsequently clustered with the anti-\kt algorithm with a distance parameter of 0.8, and then matched to the CHS AK8 jets described above used for the kinematic selections.

Subsequently, the soft-drop algorithm~\cite{Dasgupta:2013ihk,Larkoski:2014wba}, designed to remove contributions from soft radiation, is applied to the AK8 PUPPI jets. The soft-drop jet mass $m_{\mathrm{j}}$ is defined as the invariant mass associated with the four-momentum of the soft-drop jet. Dedicated mass corrections, derived from data in a region enriched with \ttbar events containing merged hadronic {\PW} decays, are applied to $m_{\mathrm{j}}$ to remove any dependence on jet \pt, and to match the jet mass and resolution observed in data. After the application of these corrections, the {\PW}(\qqbar) jet mass resolution is measured to be 10\% in the jet \pt range considered.

The two-prong hadronic decays of {\PW} and \PZ boson candidates are used to discriminate against jets
initiated from single quarks and gluons. The constituents of the jet are clustered again using the \kt algorithm~\cite{Cacciari:2008gp}, and the procedure is stopped when N subjets are obtained. Subsequently, the N-subjettiness as defined in Ref.~\cite{Thaler:2010tr} is calculated on the PUPPI-corrected jet for the one and two-subjet hypotheses as
\begin{equation}
\tau_{\mathrm{N}} = \frac{1}{d_0} \sum_k p_{\text{T},k} \min( \Delta R_{1,k}, \Delta R_{2,k}, \dots, \Delta R_{\mathrm{N},k}).
\end{equation}
The normalization factor is given by the factor $d_0 = \sum_k p_{\text{T}, k}\,R_0$, where $R_0$ is the radius of the original jet, the index $k$ increments over the jet constituents, and $\Delta R_{J,k} = \sqrt{\smash[b]{(\Delta\eta_{J,k})^2+(\Delta\phi_{J,k})^2}}$ are the angular distances in terms of the differences in $\eta$ and azimuth ($\phi$) calculated between the $k$th  jet constituent and the axes of the $J$th subjet.
Small values of the ratio of 2-subjettiness to 1-subjettiness, $\tau_{21} = \tau_2 / \tau_1$, correspond to a high compatibility with the hypothesis that the jet is produced by two partons from the decay of a massive object, rather than arising from a single parton. The efficiency of the $\tau_{21}$ selection is measured from data in a \ttbar-enriched sample~\cite{CMS-PAS-JME-16-003}.

Jets originating from the dominant $\bbbar$ decays of Higgs bosons are likely to have two displaced vertices because of the long lifetime and large mass of the \cPqb\ quarks. Following the procedure above, jet clustering using the anti-\kt algorithm is halted when two subjets are identified. The inclusive, combined secondary-vertex \cPqb\ tagging algorithm~\cite{Sirunyan:2017ezt} is applied to the two subjets, which are considered as \cPqb-tagged if they pass a working point that provides a misidentification rate of $\approx$10\% while maintaining an $85\%$ efficiency. To remove backgrounds containing top quark decays, events with AK4 jets that do not overlap with the AK8 jet are subjected to a veto based on the same \cPqb\ tagging algorithm, but with a working point corresponding to an efficiency of $\approx$70\% for identifying \PB\ hadrons and a $\approx$1\% misidentification rate. The ratio of the \cPqb\ tagging efficiency determined from data and simulation is used as a scale factor to correct the simulated events.

A dedicated algorithm is used to reconstruct Higgs bosons decaying to one or two $\tauh$ candidates~\cite{CMS-DP-2016-038}.
The procedure begins by using the Cambridge--Aachen algorithm~\cite{Wobisch:1998wt} with
a distance parameter of 0.8 to identify jets with a large cone size,
called CA8 jets.
For each CA8 jet with $\pt>100\GeV$, the last step of the clustering is retracted, obtaining two subjets. If these subjets are found to have $\pt>10\GeV$ and satisfy the mass drop-condition, which requires $\mathrm{max}(m_{\mathrm{subjet 1}},m_{\mathrm{subjet 2}})/m_{\mathrm{CA8 jet}}<2/3$, the two subjets are used as seeds in the standard \Pgt\ lepton reconstruction and the ``hadron-plus-strips" algorithm~\cite{Chatrchyan:2012zz} is applied to them to identify $\tauh$ candidates. If the \pt and mass drop conditions are not met, the unclustering and identification procedures are repeated (iteratively) for the most energetic subjet.
The $\tauh$ candidates selected through the hadron-plus-strips algorithm are then required to have $\abs{\eta}<2.3$ and $\pt>20\GeV$, and to satisfy a multivariate-discriminator threshold, obtained using a boosted decision tree technique~\cite{BDT}. This algorithm is trained to discriminate between genuine $\tauh$ and generic jets, using variables related to energy deposits and track impact parameters that are correlated with the \Pgt\ lepton lifetime.
If no $\tauh$ candidates are identified with this method, then the procedure is repeated using AK4 jets as seeds, with similar selection requirements.
The $\tauh$ candidate of highest \pt is required to satisfy an isolation requirement that corresponds to a 50--60\% efficiency in the considered topology. If two $\tauh$ candidates are identified, as in the $\tauh \tauh$ channel, the isolation requirement on the $\tauh$ of second highest \pt is relaxed to achieve a 70--80\% efficiency. The probability to misidentify a CA8 jet as an $\PH\to\TT$ decay is below 0.1\%, after these selection criteria.

Electrons are reconstructed in the region $\abs{\eta}<2.5$ by matching energy deposits in the ECAL with tracks reconstructed in the tracker~\cite{Khachatryan:2015hwa}. Electron identification is based on the distribution of energy deposited along the electron trajectory and the direction and momentum of the track in the inner tracker.
Additional requirements are applied to remove electrons produced through photon conversions. Electrons are also required to be isolated from other particles in the detector, by imposing an upper threshold on the isolation parameter. The electron isolation parameter is defined as the magnitude of the $\pt$ sum of all the PF candidates (excluding the electron) within $\Delta R< 0.3$ around the electron direction, after the contributions from pileup and particles associated with reconstructed $\tauh$ candidates within the isolation cone are removed.

Muons are reconstructed within the acceptance of the CMS muon system, $\abs{\eta}<2.4$, using information from both the muon spectrometer and the silicon tracker~\cite{Chatrchyan:2012xi}. Muon candidates are identified based on the compatibility of tracks reconstructed in the silicon tracker with tracks reconstructed from a combination of hits in both the tracker and the muon detector. In addition, the trajectory is required to be compatible with originating from the primary vertex, and to have a sufficient number of hits in the tracker and muon systems. Muons are required to be isolated by imposing a limit on the magnitude of the $\pt$ sum of all the PF candidates (excluding the muon) within $\Delta R<0.4$ around the muon direction, after the contributions particles associated with reconstructed $\tauh$ candidates within the isolation cone are removed.

The missing transverse momentum vector $\ptvecmiss$ is defined as the negative vectorial sum of the momenta of all PF candidates associated with the primary vertex projected onto the plane perpendicular to the beam direction. The missing transverse momentum $\ptmiss$ is defined as the magnitude of $\ptvecmiss$.
The observable \mht is defined as the magnitude of the vectorial \pt sum of all AK4 jets with $\pt>30\GeV$ and $\abs{\eta}<3.0$.

\section{Event selection}\label{sec:eventSelection}

Events are selected using a set of triggers that require \ptmiss or \mht larger than 90\GeV, in combination with additional trigger requirements, such as the presence of a jet with $\pt>80\GeV$.
The efficiency of the trigger for events subsequently satisfying the offline event selection, measured in an independent sample of events selected with muon triggers, is verified to be $>95$\%, with an uncertainty of 2\%.

All events are required to contain one Higgs boson candidate decaying to $\ell\tauh$ or $\tauh\tauh$.
The other boson candidate is reconstructed as a jet, using the same kinematic criteria in all categories.
Its soft-drop jet mass must be in the interval of 65--135\GeV. If the mass is in the range 65--85\GeV, the candidate is classified as a {\PW} boson, if it is in the range 85--105\GeV it is classified as a \PZ boson, and if it is the range 105--135\GeV it is considered to be a Higgs boson. To discriminate against backgrounds, the jets are required to have small values of $\tau_{21}$, and events are divided into categories of high-purity (HP) if $\tau_{21}<0.4$, and low-purity (LP) if $0.4<\tau_{21}<0.75$. A jet is \PV tagged if it fulfills the soft-drop jet mass and $\tau_{21}$ requirements.
The normalization scale factors $0.99\pm0.06$ for the HP category and $0.96\pm0.11$ for the LP category~\cite{CMS-PAS-JME-16-003} are applied to simulated events with genuine hadronic boson decays. Higgs boson jet candidates are classified according to the number of subjets (1 or 2) that pass the \cPqb\ tagging selection. Subjet \cPqb\ tagging is not used for jets compatible with {\PW} or \PZ candidates and no N-subjettiness requirement is applied to the Higgs boson candidate jet.
If neither the N-subjettiness nor the \cPqb\ tagging requirements are satisfied, the event is discarded.

Events are divided into categories depending on the number of identified $\tauh$ candidates (1 or 2), and on the classification of the large jet cone, \ie either HP or LP in $\tau_{21}$, or either 1 or 2 \cPqb-tagged subjets.

Since the undetected neutrinos carry a significant fraction of the momentum in the \Pgt\Pgt\ system, signal events are expected to have a large \ptvecmiss, thereby justifying the use of triggers that require large \ptmiss or \mht. A stringent offline requirement of greater than 200\GeV is applied to the reconstructed \ptmiss, to ensure a stable trigger efficiency and to suppress the background contribution from multijet events. Events with top quark pairs or single top quarks are suppressed by removing events in which any AK4 jet not overlapping with the AK8 jet is \cPqb-tagged.

Several selection requirements are applied to remove SM backgrounds, such as meson and baryon resonances, {\PZ}+jets, {\PW}+jets and \ttbar and single top quark production. The angular distance $\Delta R_{\Pgt\Pgt}$ should be smaller than 1.5, in order to reject {\PW}+jets events in which a jet misidentified as a \Pgt\ lepton is typically spatially well-separated from the genuine lepton. To further increase the signal purity, the di-tau mass, as estimated from the \textsc{SVfit} procedure~\cite{SVFIT, 1742-6596-513-2-022035, Bianchini:2016yrt}, should be between 50 and 150\GeV. The \textsc{SVfit} algorithm, based on a likelihood approach, estimates the di-$\tau$ system mass using the measured momenta of the visible decay products of both $\tau$ leptons, the reconstructed \ptvecmiss, and the \ptvecmiss resolution.

Finally, the resonance candidate mass $m_\text{X}$, defined as the invariant mass of the $\PH\to\tau\tau$ candidate and the hadronically decaying boson jet, is required to be larger than 750\GeV in order to ensure full trigger and reconstruction efficiencies.

After these selection requirements, the selection efficiency of a radion signal in the 1 and 2 \cPqb\ tag categories is 1--6\% in the $\tauh\tauh$ channel, and 3--10\% in the $\ell\tauh$ channel, for low and high resonance masses, respectively. The efficiencies for a \PWpr signal passing the \PV-tagging selection are 2--9\% and 8--19\% in the $\tauh\tauh$ and $\ell\tauh$ channels, respectively. Events in which $\PH\to\PV\PV$ are found contribute up to an additional 20--30\% of the total signal yield in the $\ell\tauh$ channels, and less than 10\% in the $\tauh\tauh$ channels.

\section{Background estimation}
\label{sec:backgroundEstimation}

The main sources of background originate from top quark pair production and from the production of a vector boson in association with jets ({\PZ}+jets and {\PW}+jets), while minor contributions arise from single top quark, diboson, and multijet production. These background contributions are split into either \ttbar and single top quark production (\ttbar, \cPqt), or into \Vjets production. The latter includes {\PZ}+jets and {\PW}+jets, multijet, and SM diboson production.

The shape of the distribution of the top quark pair and single top quark background is determined from simulation, while the normalization is determined from data through dedicated control regions that are enriched in top quark events. Control regions having a purity larger than 80\% for top quarks are selected by inverting the \cPqb\ tag veto on the AK4 jets and tightening the \cPqb\ tagging criteria. Events are separated according to the requirements of large-cone jet identification. Data are found to be well-described by simulations in terms of the jet and dijet resonance-mass distributions. Multiplicative scale factors are used to correct for the difference in normalization of data and simulation in the control regions, after subtracting the other background contributions. Scale factors obtained in control regions of $\ell\tauh$ events are applied also to the $\tauh\tauh$ channels, where there are fewer events.
The normalization of top quark production processes in each region is also corrected using the scale factors reported in Table~\ref{tab:topSF}.

\begin{table}[ht]
\centering
\topcaption{Normalization scale factors for top quark production for different event categories, depending on the \PV tagging and \PH tagging requirement applied. Uncertainties are due to the limited number of events in the control regions and the uncertainty in the \cPqb\ tagging efficiency.}\label{tab:topSF}

\begin{tabular}{ccccc}

Channel  & $\tau_{21}$ LP &  $\tau_{21}$ HP & 1 \cPqb-tagged subjet & 2 \cPqb-tagged subjets \\\hline
 $\ell\tauh$ & 0.96 $\pm$ 0.04  & 1.06 $\pm$ 0.06 & 1.00 $\pm$ 0.06 & 1.11 $\pm$ 0.15 \\
\end{tabular}

\end{table}

\begin{figure}[!htb]
  \centering
  \includegraphics[width=.495\textwidth]{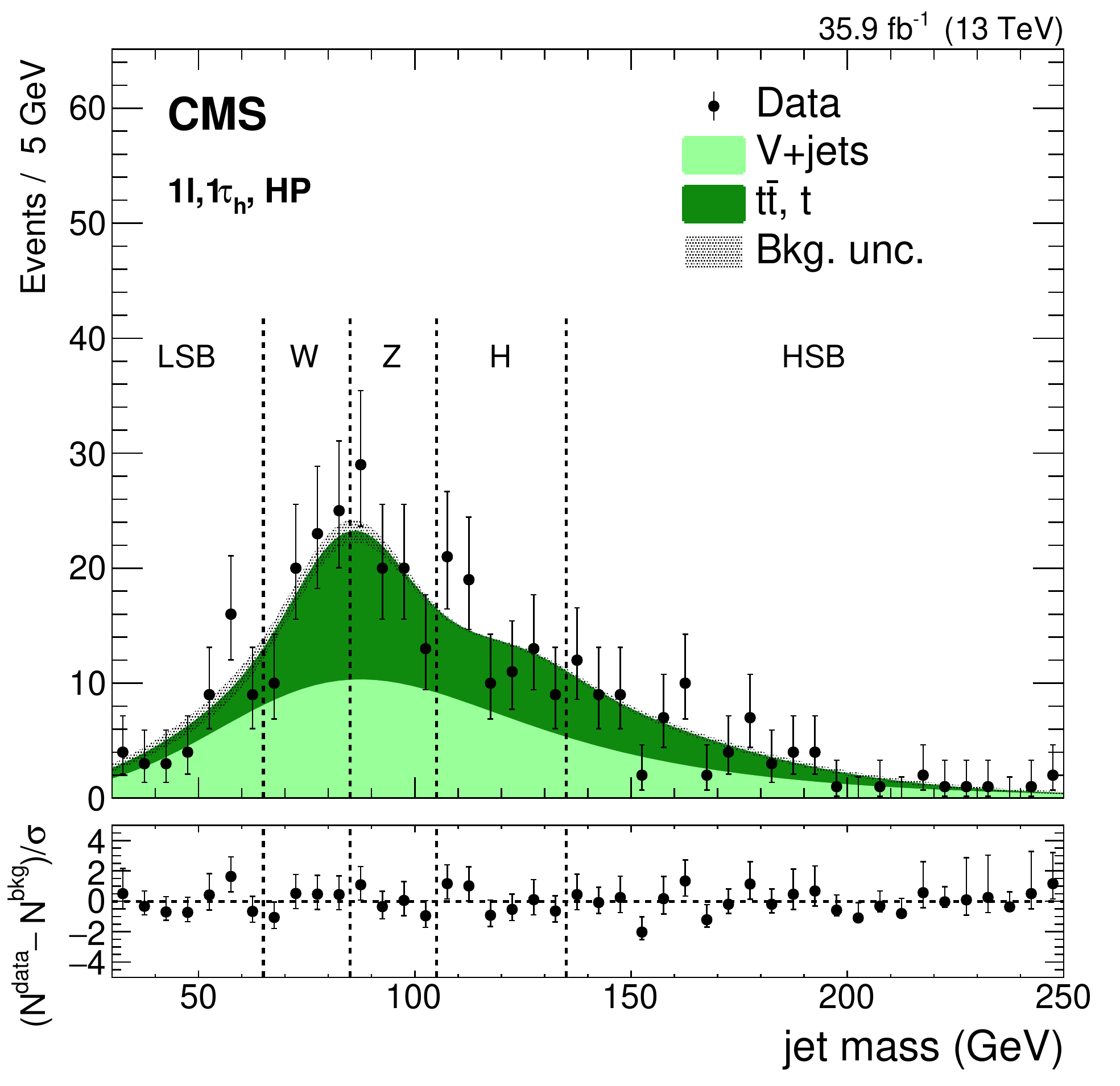}
  \includegraphics[width=.495\textwidth]{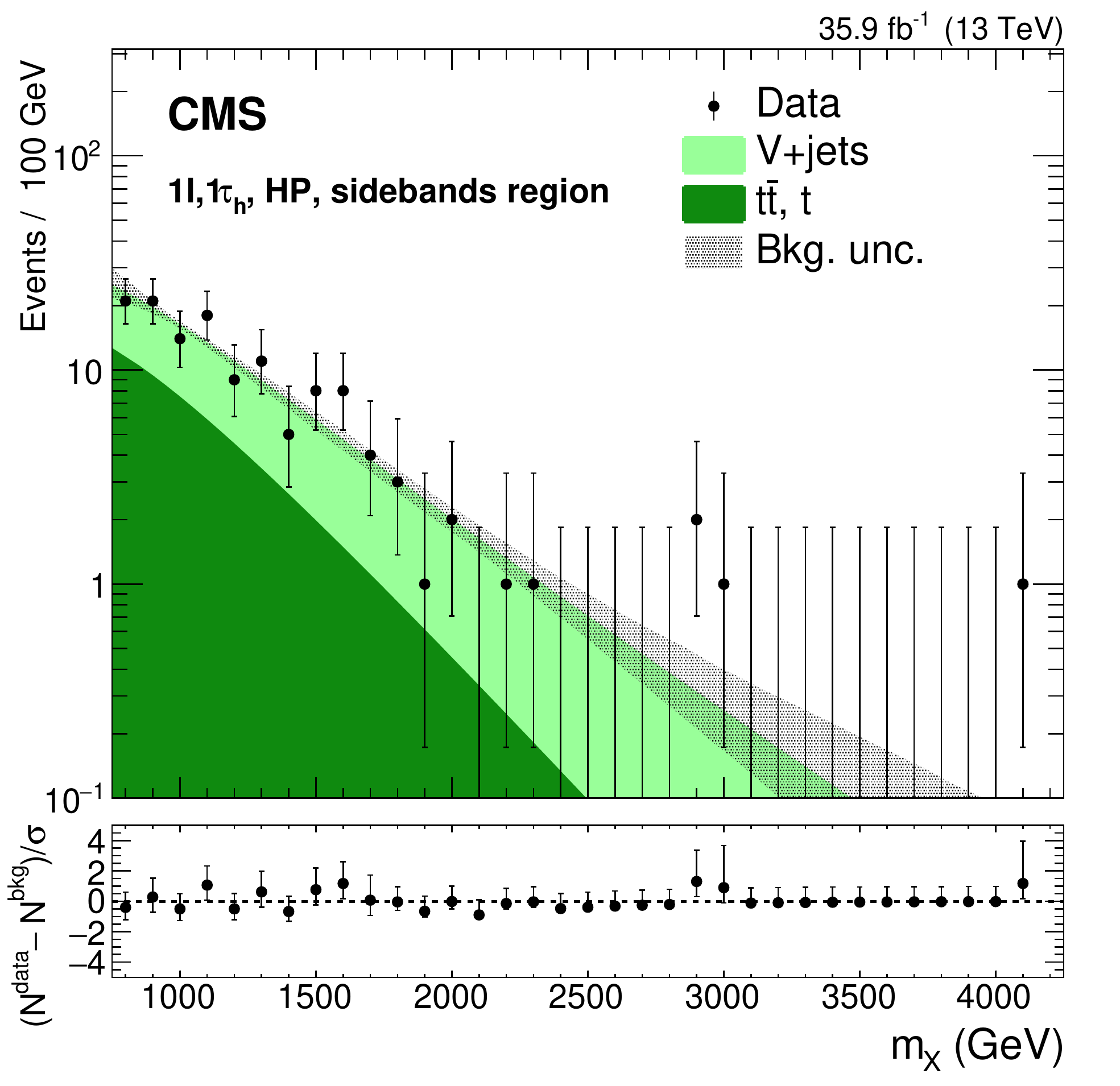}

  \caption{Soft-drop jet mass distribution in data in the HP $\ell\tauh$ category, together with the background prediction (fitted to the data as explained in the text)(\cmsLeft). Spectrum of the resonance mass in data events in the SBs (\cmsRight) used for the estimation of the \Vjets distribution in the SR. The lower panels depict the pulls in each
bin, $(\mathrm{N}_{\mathrm{data}} - \mathrm{N}_{\mathrm{bkg}})/\sigma$, where $\sigma$ is the statistical uncertainty in data, as given by the Garwood interval \cite{Garwood}.}
  \label{fig:XWhmnqq_Wwind_Alpha}
\end{figure}

The estimated contribution from \Vjets backgrounds is based on data, in regions defined by applying the complete signal selection apart from the jet-mass requirements.
Data are divided into the $\tauh\tauh$ and the $\ell\tauh$ channels. Two jet-mass sidebands (SB) are defined with jet masses in the range of 30--65\GeV for the low sideband (LSB), or above 135\GeV for the high sideband (HSB), and used to predict the background contribution in the signal region (SR). Analytic functions are fitted to the simulated distributions of the jet mass, separately for \Vjets and for top quark processes. The former is modeled by a third-order polynomial or an exponential convoluted with an error function, and the latter with a sum of one or two Gaussian functions to model the {\PW} and top quark distributions.
The background model, composed of the sum of the functions modeling the two backgrounds, is fitted to the data in the jet mass sidebands. In this unbinned likelihood fit, only the \Vjets distribution shape and normalization parameters are left free to float. The number of expected events in the SR is then obtained by integrating the background components in the jet mass window compatible with the signal hypothesis. An indicative fit is shown in Fig.~\ref{fig:XWhmnqq_Wwind_Alpha} (\cmsLeft) for the HP $\tau_{21}$ category of $\ell\tauh$ events.
The procedure is repeated with an alternative function used for modeling the \Vjets jet mass, with the observed difference in the normalization taken to be the associated systematic uncertainty.
The expected number of background events in each signal region is reported in Table~\ref{tab:ExpectedBkgSR}.

The distribution of the \Vjets background resonance mass ($m_\text{X}$) in the SR is estimated from the SBs through a transfer function $\alpha(m_\text{X})$, which accounts for the small kinematical differences and the correlations involved in the interpolation from the SBs to the SR, and does not depend on the systematic uncertainties that affect the simulated \Vjets spectra, since they cancel out in the ratio. The $\alpha$ function is determined from simulations as
\begin{equation}
\alpha(m_{\mathrm{X}}) = \frac{\mathrm{N}^{\mathrm{MC},\Vjets}_{\mathrm{SR}}(m_\text{X})}{\mathrm{N}^{\mathrm{MC},\Vjets}_{\mathrm{SB}}(m_\text{X})},
\end{equation}
where $\mathrm{N}^{\mathrm{MC},\Vjets}_{\mathrm{SR}}(m_\text{X})$ and $\mathrm{N}^{\mathrm{MC},\Vjets}_{\mathrm{SB}}(m_{\mathrm{X}})$ are analytic functions fitted to the simulated $m_\mathrm{X}$ distributions in the SR and SB regions, respectively. Depending on the category, the function can either be an exponential using one or two parameters, or a power law with one parameter.
The distribution of the \Vjets background in the SR is then estimated by fitting an analytic function to data in the SBs, after subtracting the top quark background estimated from simulation, and multiplying by the $\alpha(m_\mathrm{X})$ transfer function. The normalization of the \Vjets is determined from the fit to the jet mass, as reported in Table~\ref{tab:ExpectedBkgSR}. The resonance mass distribution is shown in Fig.~\ref{fig:XWhmnqq_Wwind_Alpha} (\cmsRight) for $\ell\tauh$ events in the HP category.

\begin{table}[!htb]
  \begin{center}
    \topcaption{Predicted number of background events and the observed number in the signal region, for all event categories. The regions denoted by {\PW}, \PZ and \PH are intervals in the jet soft-drop mass distribution that range from 65 to 85\GeV, from 85 to 105\GeV, and from 105 to 135\GeV, respectively. Separate sources of uncertainty in the expected number are reported as the statistical uncertainty in the \Vjets contribution from the fitting procedure (fit), the difference between the nominal and alternative function form chosen for the fit (alt), and the uncertainty in the background from top quarks from the fit to the simulated jet mass spectrum.}\label{tab:ExpectedBkgSR}
\vspace{5mm}
   \begin{tabular}{ccccccc}

      \multicolumn{3}{c}{Category} & \Vjets ($\pm$ fit)($\pm$ alt) & \ttbar, \cPqt  & Total exp. events & Data \\
      \hline

      \multirow{4}{*}{{\PW} region} & \multirow{2}{*}{HP} & $\ell\tauh$ & $38 \pm 7 \pm 12$ & $37.8 \pm 0.6$ & $76 \pm 14$ & $78$ \\
      &  & $\tauh\tauh$ & $13.0 \pm 3.2 \pm 0.2$ & $16.0 \pm 1.8$ & $29.0 \pm 3.7$ & $45$ \\

      & \multirow{2}{*}{LP} & $\ell\tauh$ &$105.3 \pm 6.8 \pm 9.0$ & $34.2 \pm 0.9$ & $140 \pm 11$ & $120$ \\
      &  & $\tauh\tauh$ & $27.0 \pm 3.3 \pm 3.0$ & $12.3 \pm 0.6$ & $39.3 \pm 4.5$ & $37$ \\[\cmsTabSkip]

      \multirow{4}{*}{\PZ region} & \multirow{2}{*}{HP} & $\ell\tauh$ & $39.9 \pm 6.1 \pm 7.9$ & $42.4 \pm 1.0$ & $82 \pm 10$ & $82$ \\
      &  & $\tauh\tauh$ & $13.7 \pm 3.0 \pm 2.5$ & $18.0 \pm 1.8$ & $31.6 \pm 4.3$ & $33$ \\
      & \multirow{2}{*}{LP} & $\ell\tauh$ & $73.5 \pm 4.8 \pm 6.1$ & $29.1 \pm 1.9$ & $102.6 \pm 8.0$ & $92$ \\
      &  & $\tauh\tauh$ & $19.1 \pm 2.3 \pm 2.5$ & $10.4 \pm 0.8$ & $29.5 \pm 3.5$ & $33$ \\[\cmsTabSkip]

      \multirow{4}{*}{\PH region} & \multirow{2}{*}{2 {\cPqb} tag} & $\ell\tauh$ & $2.4 \pm 0.9 \pm 0.4$ & $6.9 \pm 0.6$ & $9.2 \pm 1.2$ & $10$ \\
      &  & $\tauh\tauh$ & $1.1 \pm 0.6 \pm 0.1$ & $3.8 \pm 1.8$ & $4.9 \pm 1.9$ & $5$ \\

      & \multirow{2}{*}{1 {\cPqb} tag} & $\ell\tauh$ &$29.3 \pm 3.5 \pm 6.6$ & $37.3 \pm 1.2$ & $66.6 \pm 7.5$ & $56$ \\
      &  & $\tauh\tauh$ &$11.5 \pm 2.2 \pm 2.6$ & $15.4 \pm 1.7$ & $26.9 \pm 3.8$ & $23$ \\

   \end{tabular}
 \end{center}

\end{table}

\begin{figure}[!htb]
  \centering

  \includegraphics[width=.43\textwidth]{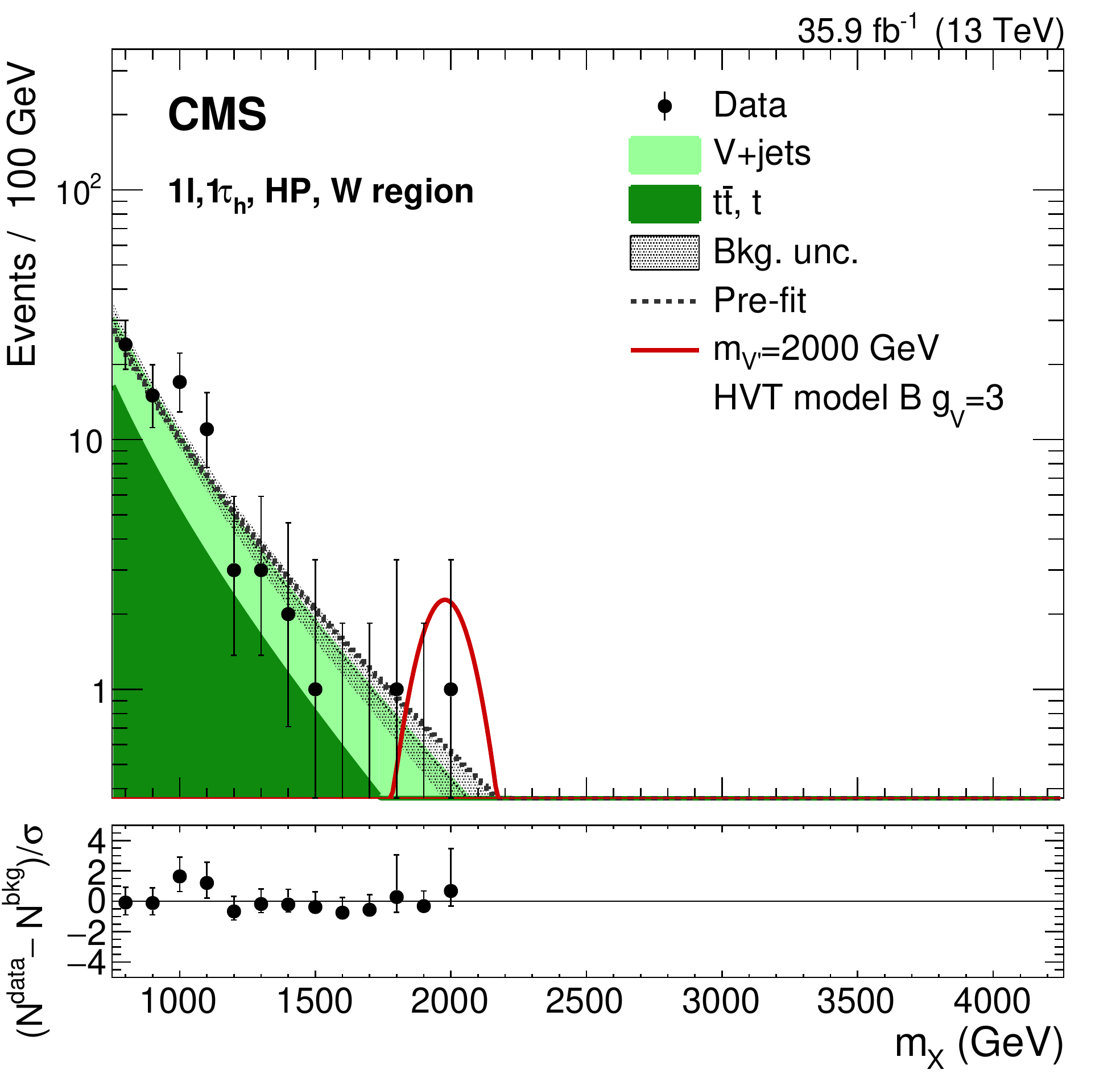}
  \includegraphics[width=.43\textwidth]{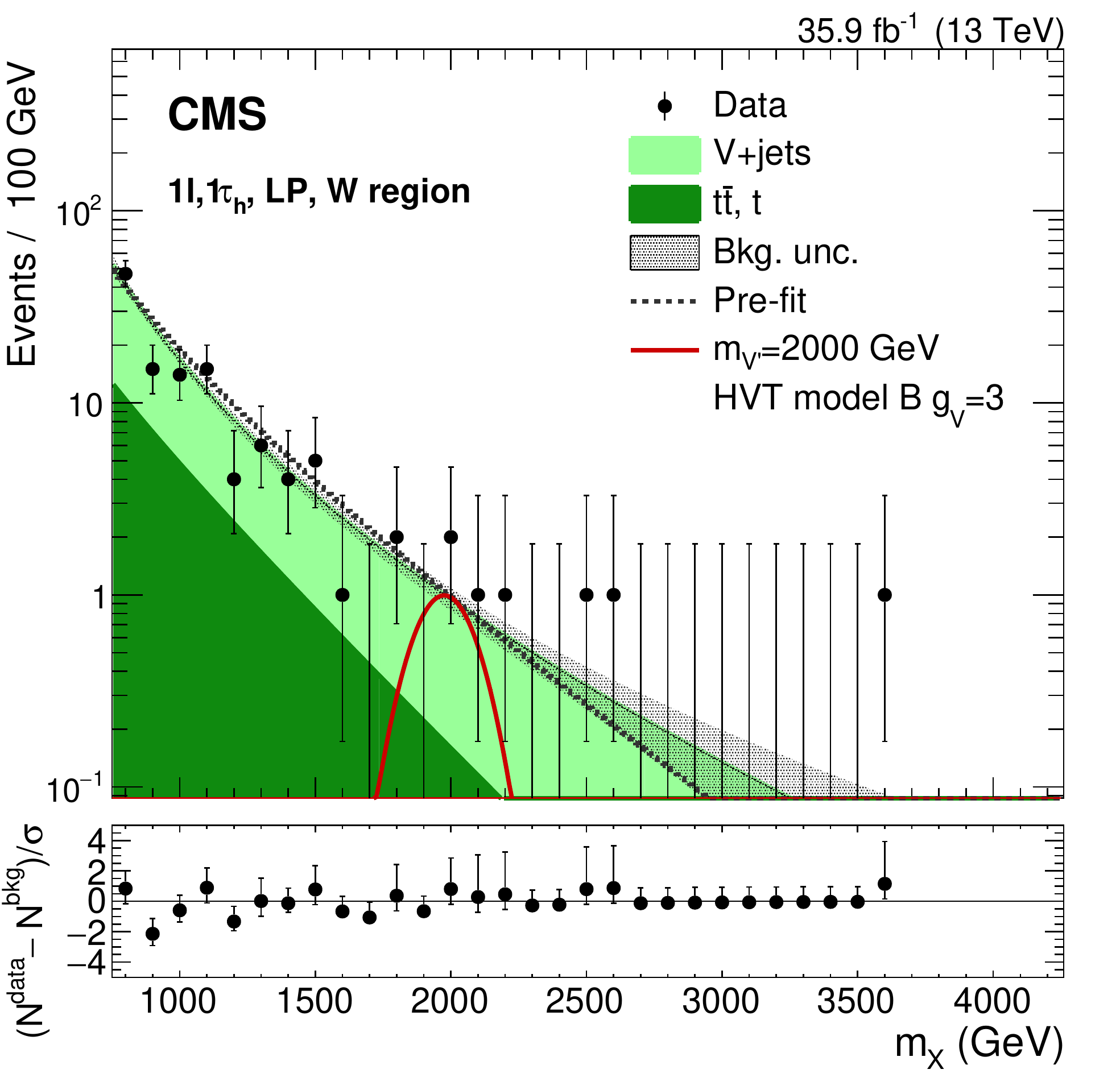}\\
  \includegraphics[width=.43\textwidth]{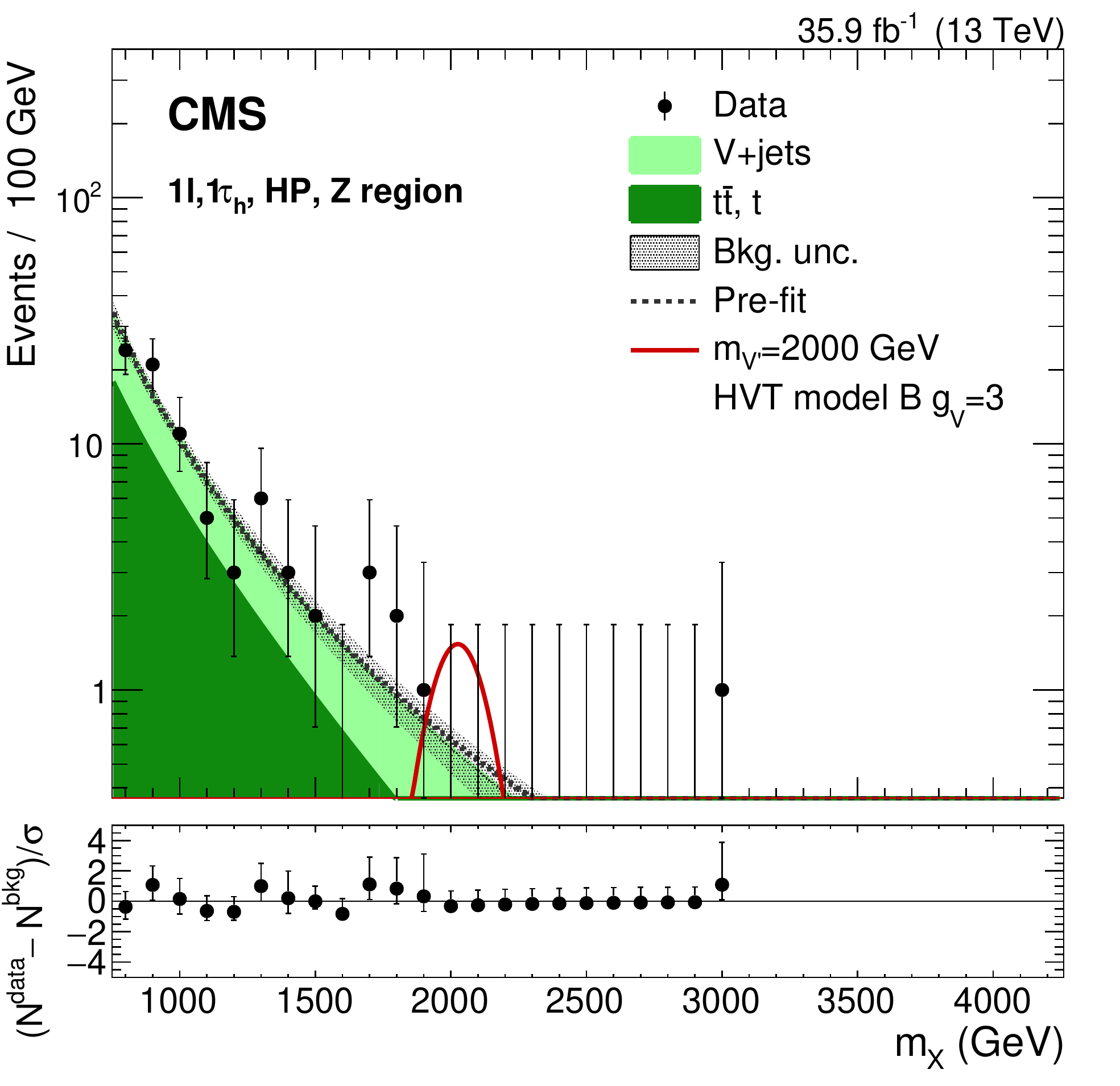}
  \includegraphics[width=.43\textwidth]{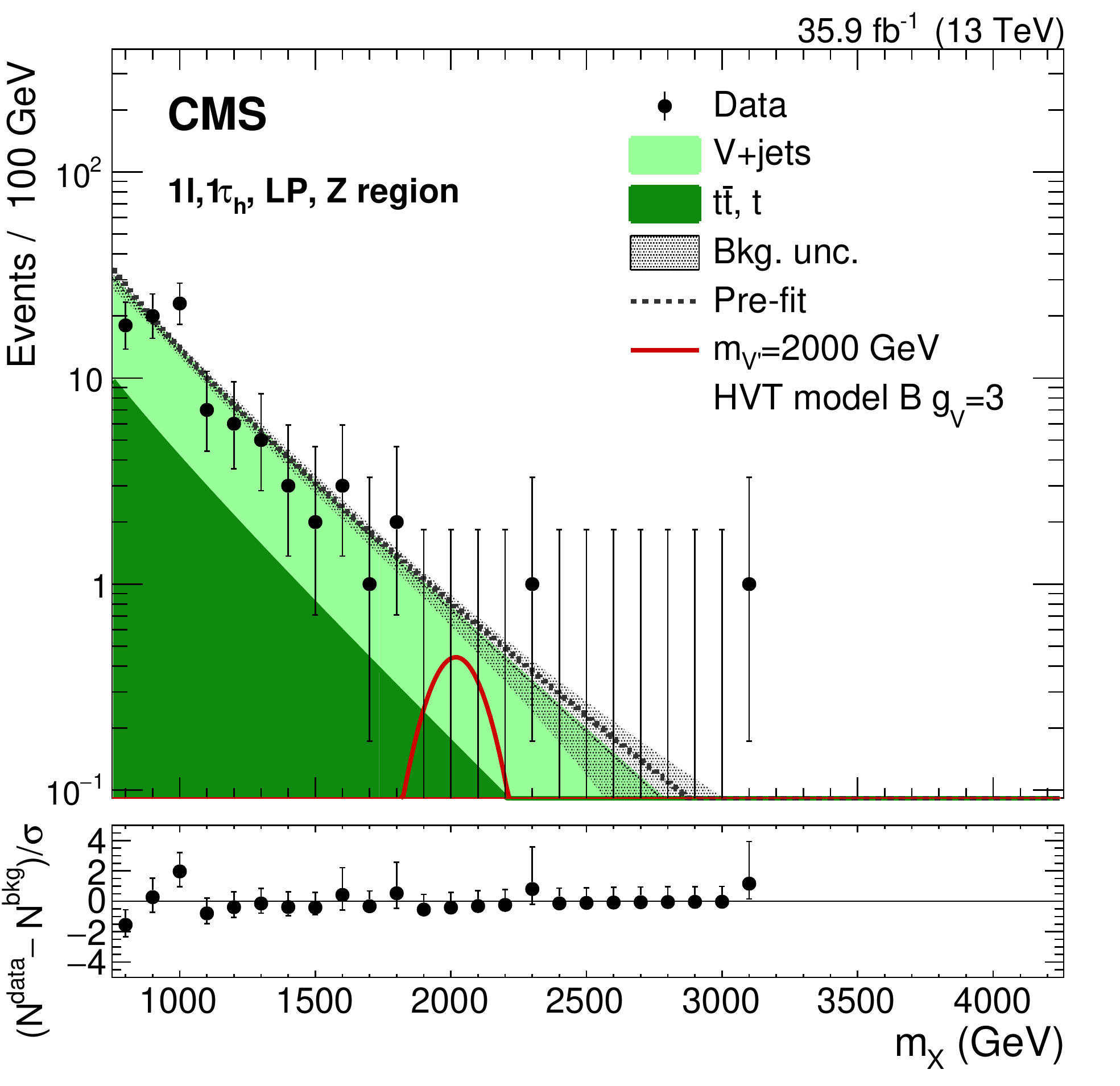}\\
  \includegraphics[width=.43\textwidth]{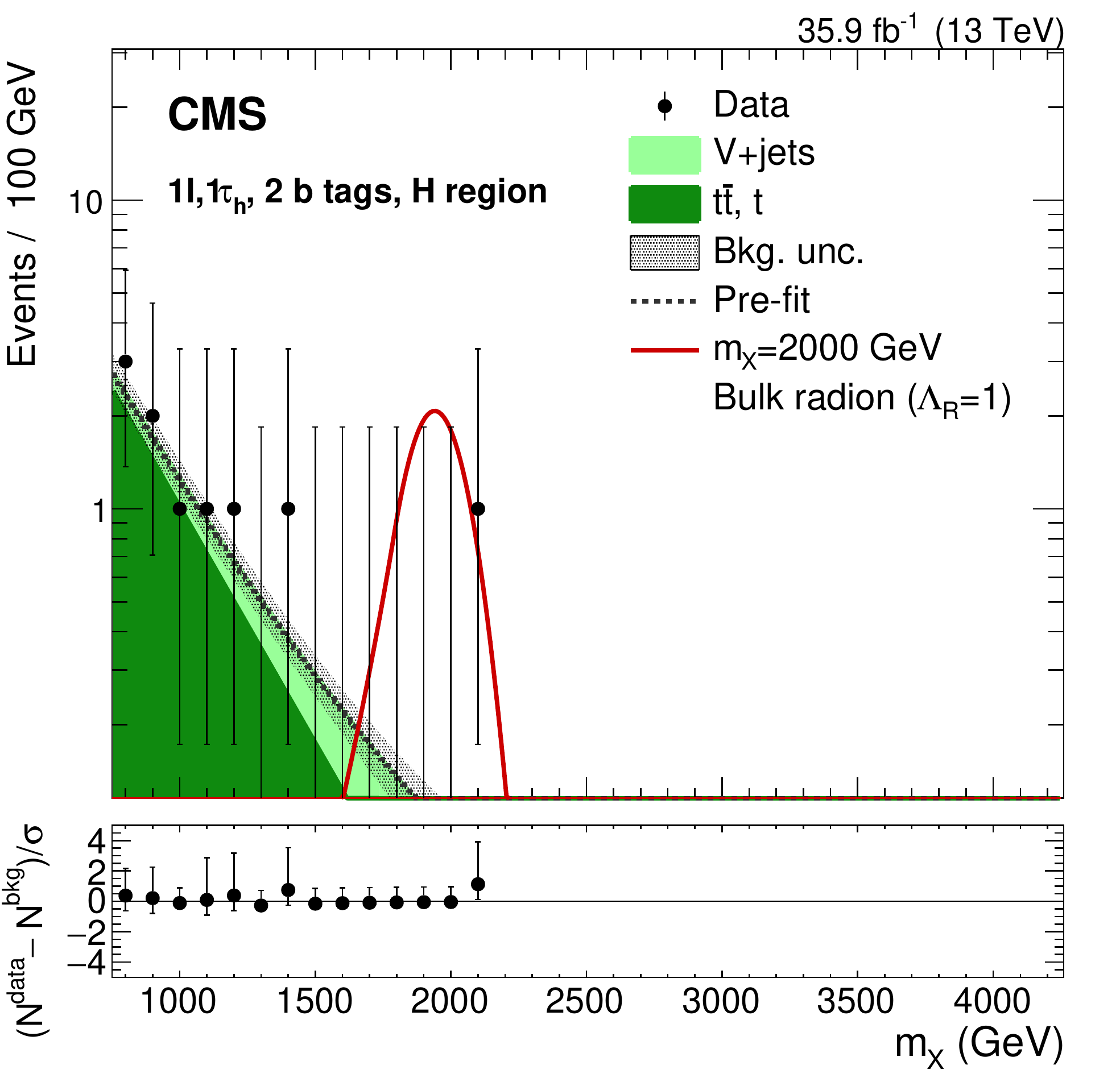}
  \includegraphics[width=.43\textwidth]{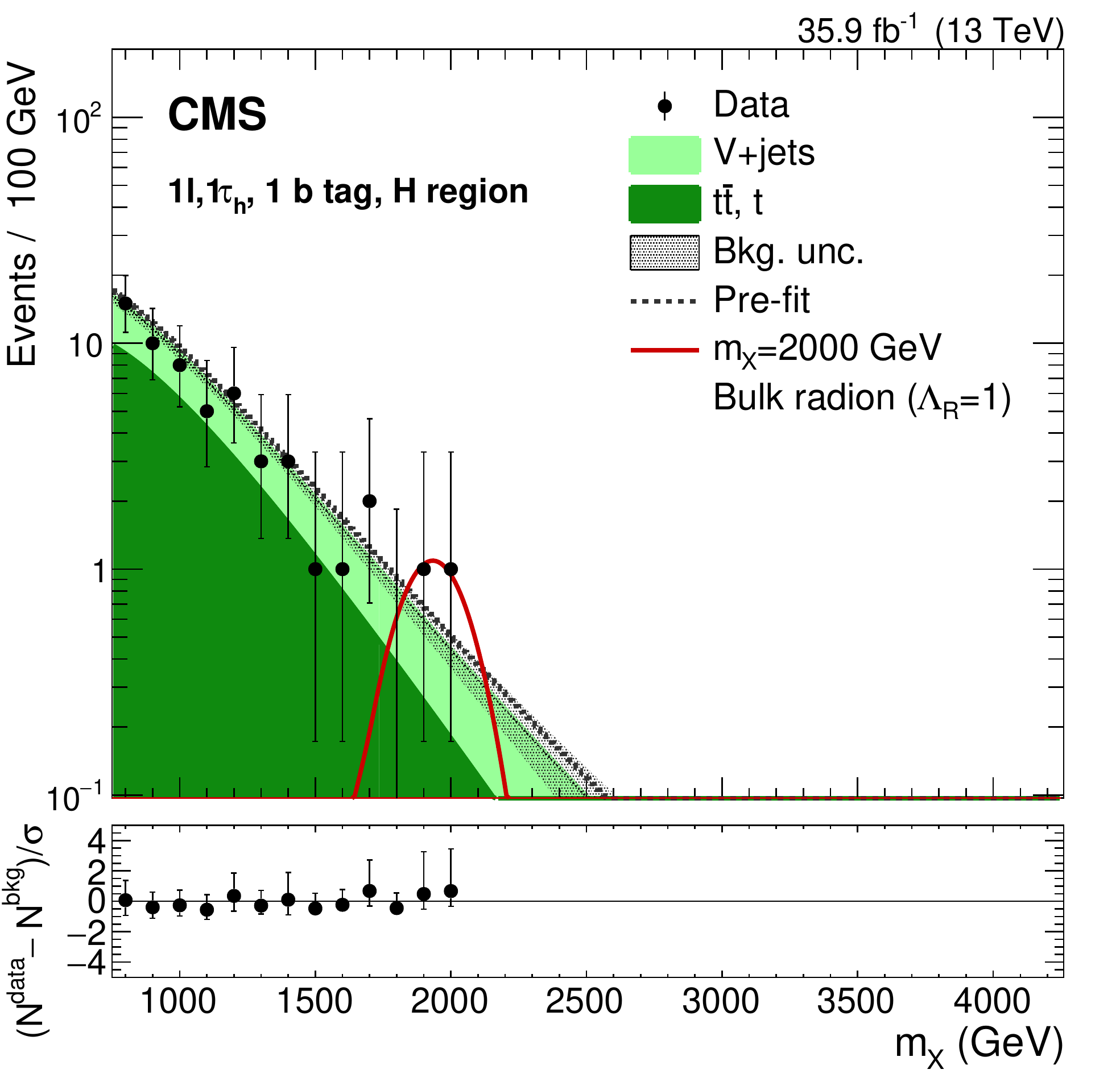}

  \caption{Data and expected backgrounds in the $\ell\tauh$ channel. The {\PW} mass window is shown in the HP (upper \cmsLeft) and LP (upper \cmsRight) categories, the \PZ mass window for the HP (middle \cmsLeft) and LP (middle \cmsRight) categories, and the \PH mass window for the two \cPqb-tagged subjet (lower \cmsLeft) and one \cPqb-tagged subjet (lower \cmsRight) categories.
The lower panels depict the pulls in each bin, $(\mathrm{N}_{\mathrm{ data}} - \mathrm{N}_{\mathrm{ bkg}})/\sigma$, where $\sigma$ is the statistical uncertainty in data, as given by the Garwood interval \cite{Garwood}, and provide estimates of the goodness of fit. Signal contributions are shown, assuming benchmark HVT model B for the \PVpr and $\Lambda_{\mathrm{R}}=1$ for the radion.}
  \label{fig:XWhmn_Exp}
\end{figure}

\begin{figure}[!htb]
  \centering
  \includegraphics[width=.43\textwidth]{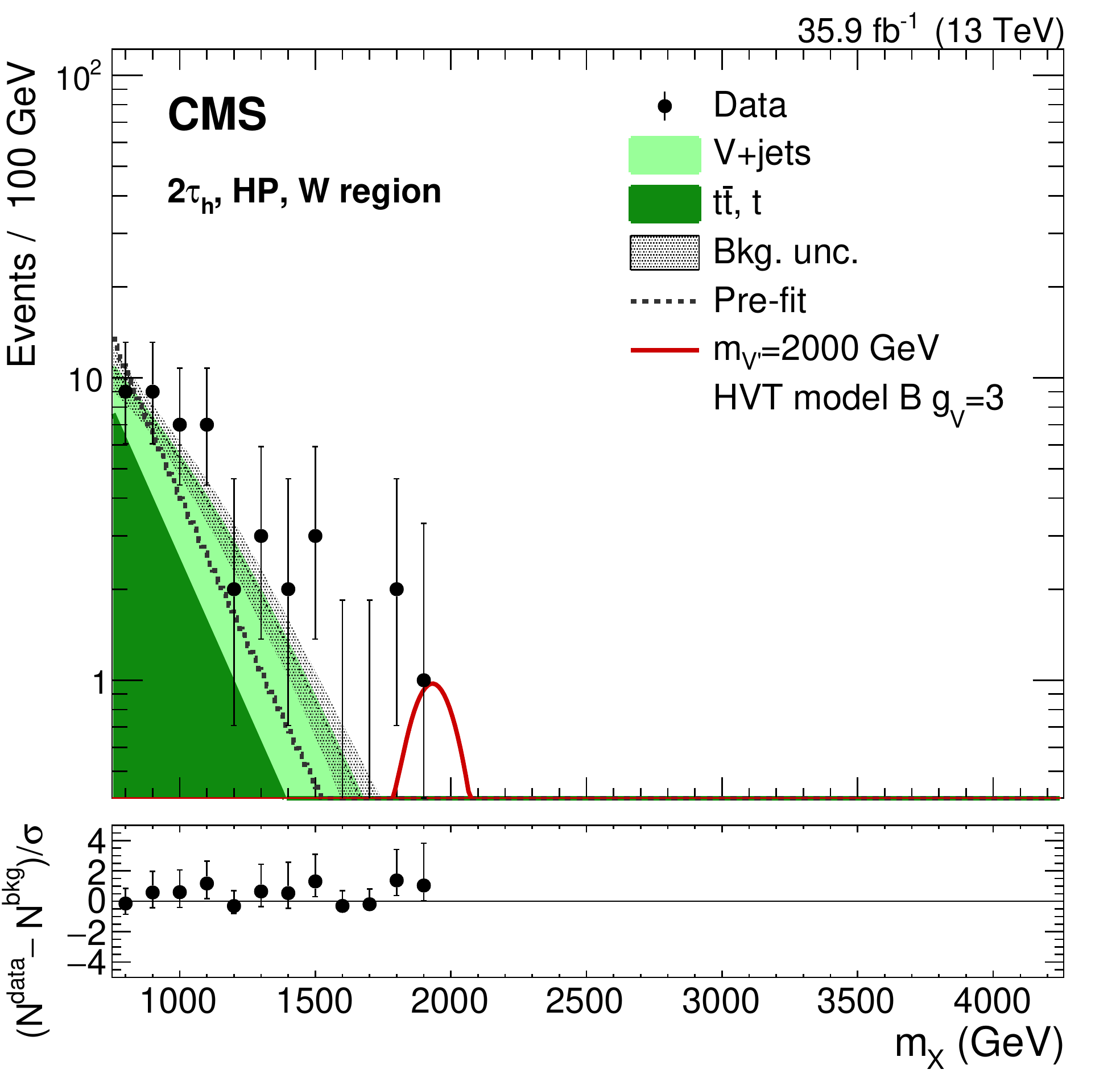}
  \includegraphics[width=.43\textwidth]{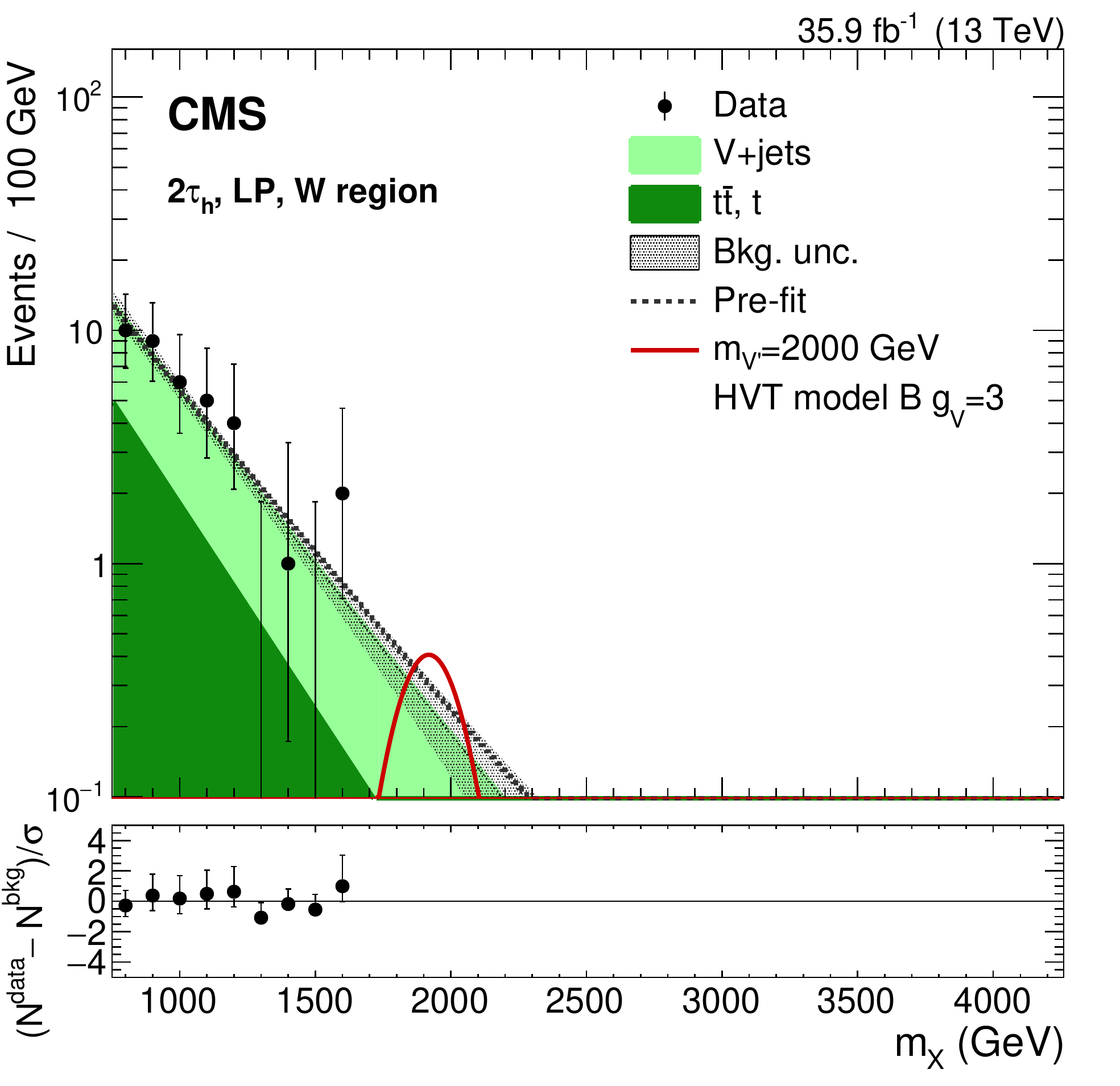}\\
  \includegraphics[width=.43\textwidth]{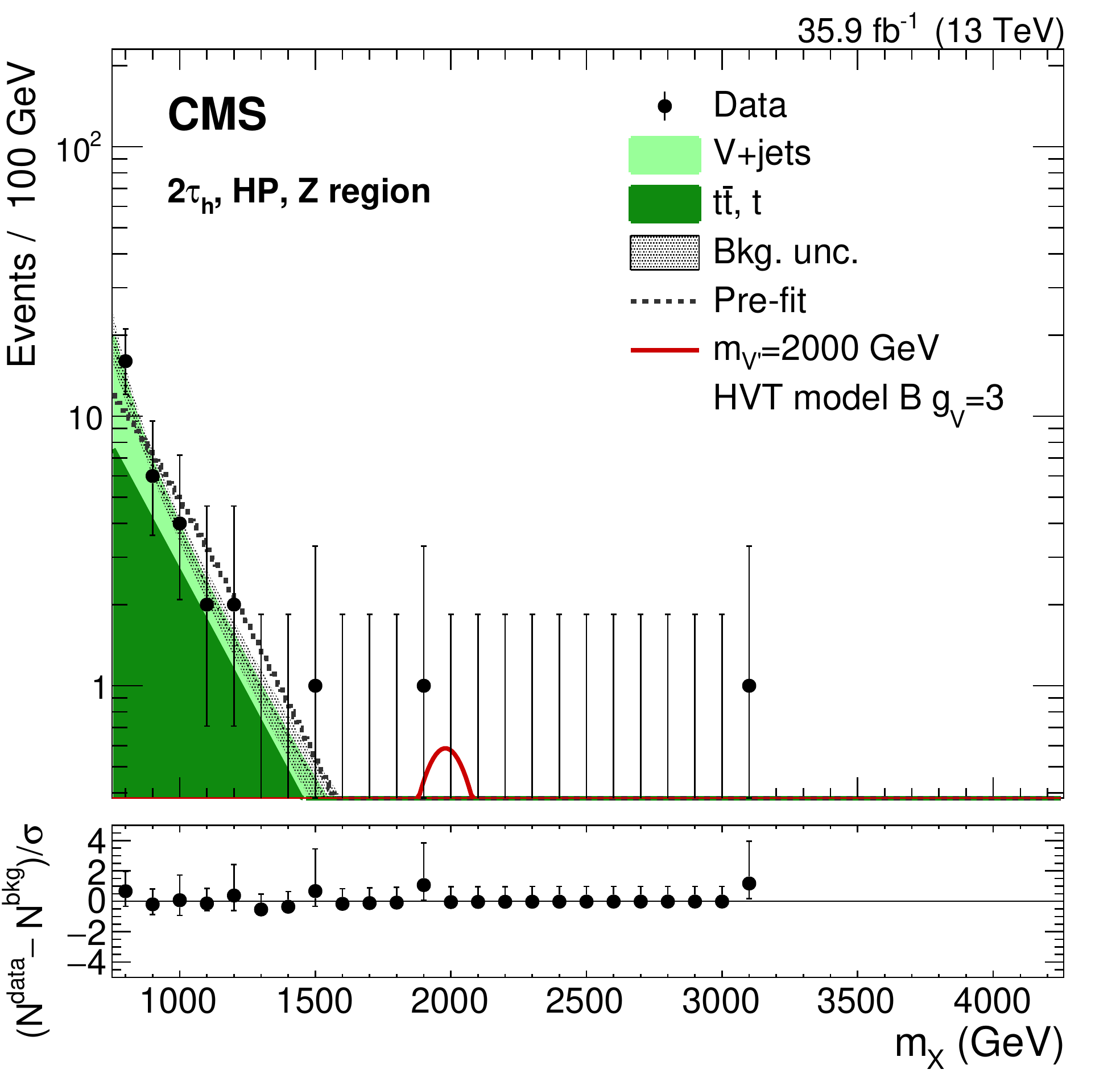}
  \includegraphics[width=.43\textwidth]{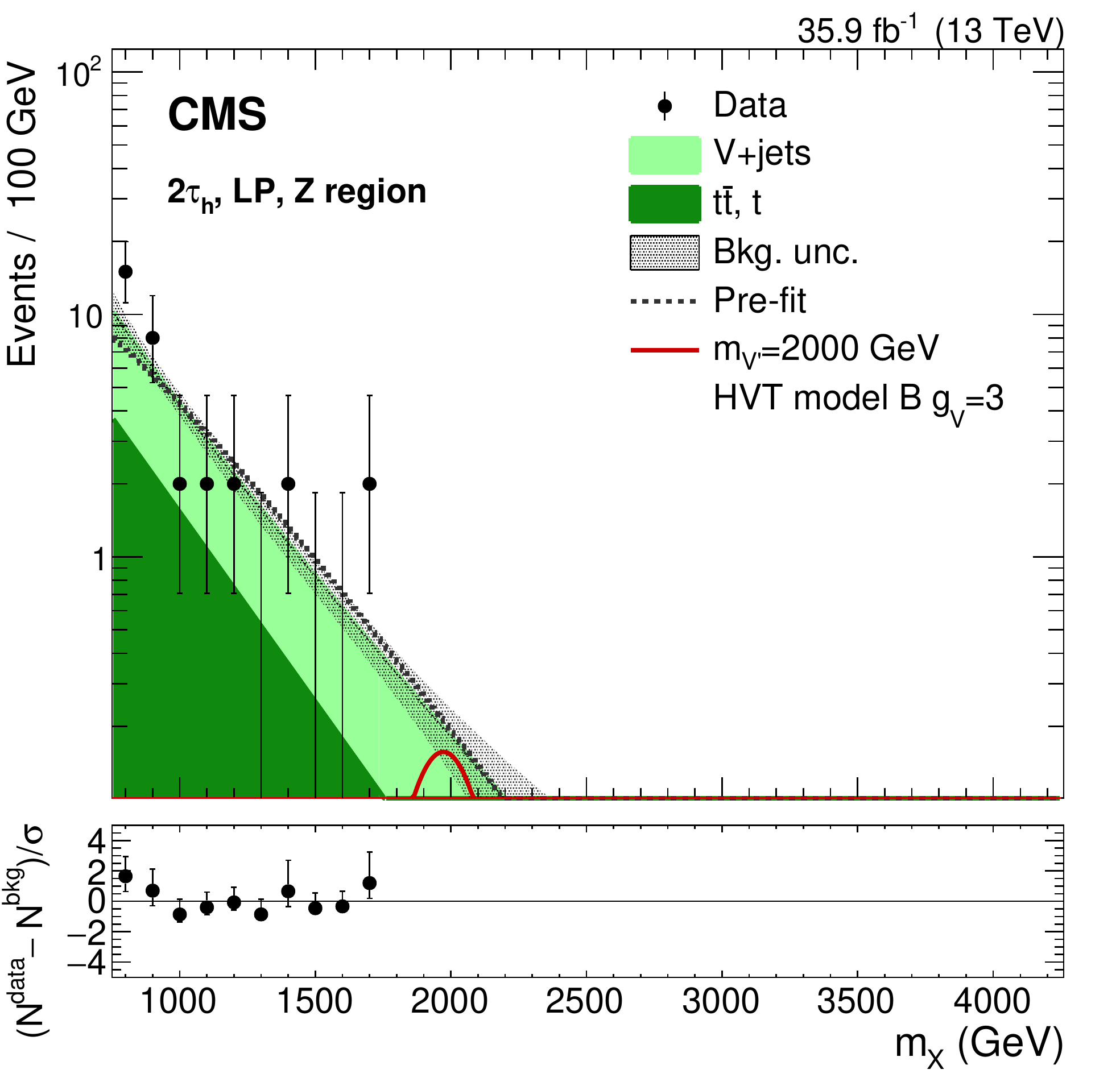}\\
  \includegraphics[width=.43\textwidth]{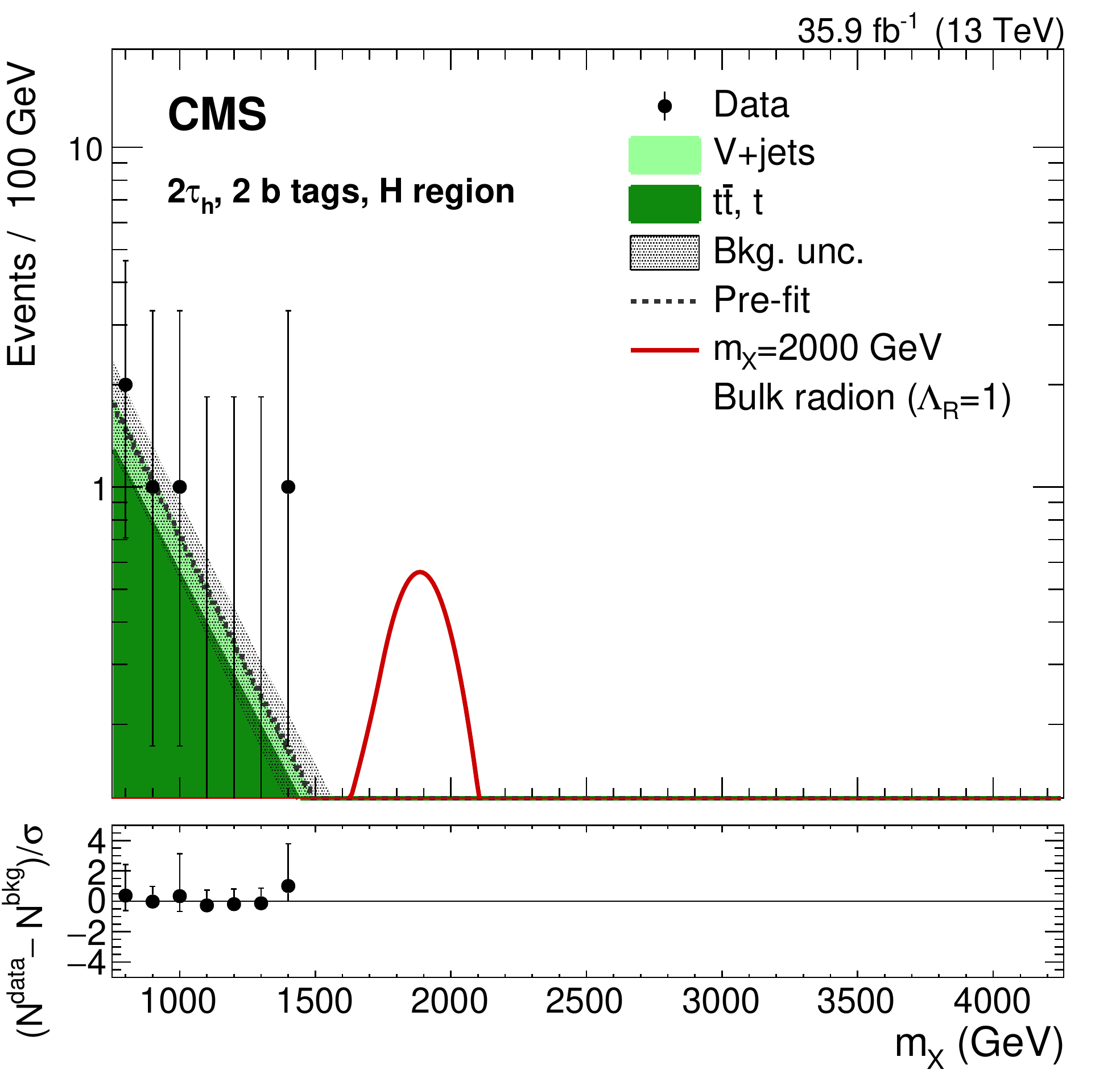}
  \includegraphics[width=.43\textwidth]{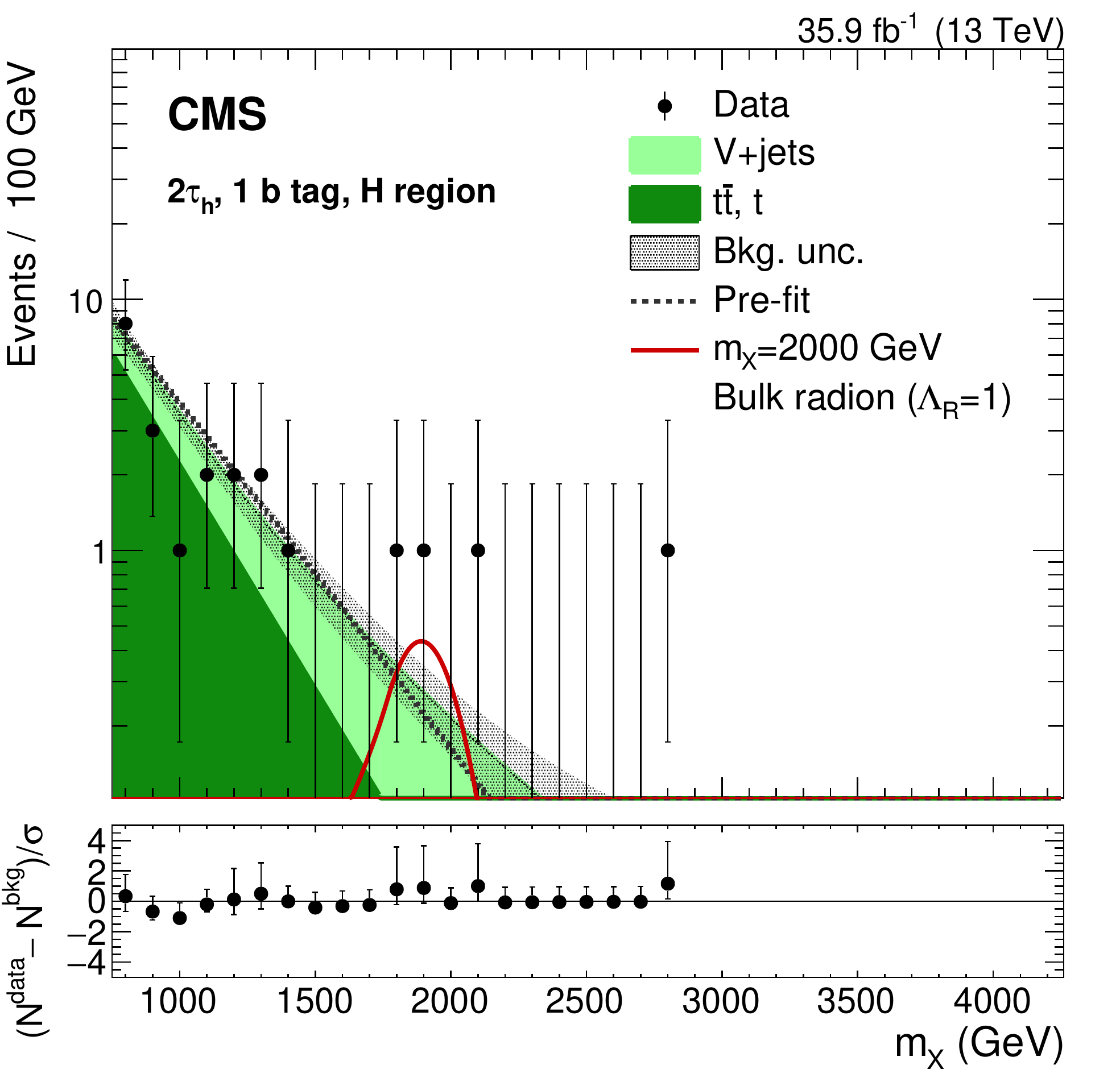}
  \caption{Data and expected backgrounds in the $\tauh\tauh$ channel. The {\PW} mass window is shown in the HP (upper \cmsLeft) and LP (upper \cmsRight) categories, the \PZ mass window for the HP (middle \cmsLeft) and LP (middle \cmsRight) categories, and the \PH mass window for the two \cPqb-tagged subjet (lower \cmsLeft) and one \cPqb-tagged subjet (lower \cmsRight) categories. The lower panels depict the pulls in each bin, $(\mathrm{N}_{\mathrm{data}} - \mathrm{N}_{\mathrm{ bkg}})/\sigma$, where $\sigma$ is the statistical uncertainty in data, as given by the Garwood interval \cite{Garwood}, and provide estimates of the goodness of fit. Signal contributions are shown, assuming benchmark HVT model B for the \PVpr and $\Lambda_{\mathrm{R}}=1$ for the radion.}
  \label{fig:Xhhen_Exp}
\end{figure}

The overall background in the SR is then expected to be:
\begin{equation}
\mathrm{N}_{\mathrm{SR}}^{\mathrm{data}}(m_{\mathrm{X}}) = \alpha(m_{\mathrm{X}}) [\mathrm{N}_{\mathrm{SB}}^{\mathrm{data}} - \mathrm{N}_{\mathrm{SB}}^{\ttbar, \cPqt}] (m_{\mathrm{X}}) + \mathrm{N}_{\mathrm{ SR}}^{\ttbar, \cPqt} (m_{\mathrm{X}}),
\end{equation}
where $\mathrm{N}_{\mathrm{SB}}^{\ttbar, \cPqt}$ and $\mathrm{N}_{\mathrm{SR}}^{\ttbar, \cPqt}$ are the distributions for the top quark process in the SB and SR, respectively.
The shape and the normalization of the distribution are fixed from simulation, with the latter corrected using the appropriate scale factors in Table~\ref{tab:topSF}.
The data in the SR and the background predictions before and after the fit in the SR are shown in Figs.~\ref{fig:XWhmn_Exp} and \ref{fig:Xhhen_Exp}.

\section{Systematic uncertainties}
\label{sec:systematicUncertainties}

Systematic uncertainties, resulting from experimental and theory sources, may affect both the normalization and shape of the signal and background distributions.

The principal background is \Vjets, and its modeling represents the largest uncertainty in the analysis.
The systematic uncertainty in the \Vjets background is dominated by the statistical uncertainty associated with the number of events in the jet mass distribution SBs in data and simulation. An additional uncertainty is related to the choice of model used for the jet mass in the \Vjets background.
It is evaluated from the differences in the expected background yields obtained when using the alternative fitting functions.
For the top quark processes, uncertainties from normalization and shape in the parametrization are propagated to the final background estimation. The single top quark and top quark pair production normalization uncertainty arises predominantly from the limited number of events in the CRs.

The uncertainties in the shape of the \Vjets distribution are estimated from the covariance matrix of the fit to $m_\text{X}$ in the sidebands and the uncertainties in the $\alpha(m_\text{X})$ ratio, which depend on the number of events in data and simulation, respectively.

The uncertainties in the trigger efficiency, and in the electron and muon reconstruction, identification, and isolation efficiencies are obtained by varying the corresponding scale factors by their uncertainty, and each is found to be 1--2\% \cite{Khachatryan:2015hwa,Chatrchyan:2012xi}.
For the $\tauh$ reconstruction and identification, the uncertainties vary between 6 and 8\% and between 10 and 13\%, depending on the resonance mass, in the $\ell\tauh$ and the $\tauh\tauh$ channels, respectively~\cite{Chatrchyan:2012zz}. A separate uncertainty due to the extrapolation of the reconstruction and identification of $\tauh$ leptons at large \pt has an impact on the signal normalization of 18\% in the $\ell\tauh$, and 30\% in the $\tauh\tauh$ channels, for a 4\TeV signal hypothesis. This uncertainty is responsible for an increase of 1\% in the width of the signal distribution.

Jet energy scale and resolution uncertainties affect both the selection efficiencies and the shape of distributions. The corrections to the jet mass scale and resolution are also taken into account, and result in a variation of 1--8\% in the expected number of signal events. The jet energy scale uncertainty accounts for a variation in signal efficiency of 1--3\%, while the variation in the jet energy resolution has an impact of 1--2\%. The effect on the mass distribution is at the level of 1--2\% for the mean and the width of the signal distribution. Event migrations between the mass windows due to the effect of jet mass scale and resolution variations are estimated to be between 2 and 15\%, depending on the signal and the vector boson mass region.

Scale factors for \PV tagging and \cPqb\ tagging represent the largest source of normalization uncertainty for the signal. Uncertainties in normalization correspond to 6 and 11\% in the HP and LP categories, respectively. An additional uncertainty from the extrapolation of the {\PW} tagging from the \ttbar scale to larger values of jet \pt is estimated using an alternative \HERWIG~\cite{Bahr:2008pv} shower model, and varies from 2 to 18\% for 0.9--4\TeV mass hypotheses and the two \PV tag categories.
In addition, the contribution to the signal normalization
uncertainty from the \cPqb\ tagging uncertainty varies between 3 (4)\% to 7 (5)\% for the 2 (1) \cPqb-tagged subjet categories.

The \Pgt\ lepton energy-scale uncertainties affect both the selection efficiencies and their distribution shapes. The effect on the signal yield amounts to 1\% in the $\ell\tauh$ channel. In the $\tauh\tauh$ channel, the effect decreases from 5\% for a resonance with mass of 0.9\TeV to 3\% for a mass of 4.0\TeV.

Normalization uncertainties from the choice of the parton distribution function (PDF) grow larger with higher resonance mass, and are larger for gluon-initiated processes than for quark-initiated processes. For \PWpr and \PZpr production, which are sensitive to quark PDFs, effects range from 6 to 37\%, while radion and graviton production depend on gluon PDFs, and result in a variation of 10 to 64\% in the number of expected signal events. Uncertainties of similar magnitude arise from factor of two variations in the factorization and renormalization scales, resulting in 3 to 13\% variations for \PWpr and \PZpr, and 10 to 19\% for radion and graviton production. While normalization uncertainties are not considered in setting limits on production, effects on signal acceptance are propagated to the final fit, amounting to 0.5--2\% for PDF uncertainties, depending on resonance mass.

Other systematic uncertainties affecting the normalization of signal and minor backgrounds considered in the analysis include pileup contributions (0.5\%) and integrated luminosity (2.5\%) \cite{CMS:lumi}. A list of the main systematic uncertainties is given in Table~\ref{tab:Sys}.

\begin{table}[!htb]
  \centering
  \caption{Summary of systematic uncertainties for the background and signal events. Uncertainties marked with ``shape'' are propagated also to the shape of the distributions, and those marked with $\dagger$ are not included in the limit bands, but instead reported in the theory band. The dash symbol is reported where the uncertainty is not applicable to a certain signal or background. The symbols $\cPq\cPq$' and $\Pg\Pg$ refer to quark-initiated and gluon-initiated processes, respectively.}
  \label{tab:Sys}
  \vspace{5mm}
  \begin{tabular}{lccccccc}

                      & \Vjets & \ttbar, \cPqt & Signal \\
    \hline
    $\alpha$-function   & shape & \NA & \NA \\
    Bkg. normalization   & 11--60\% &  2--38\% & \NA \\
    Top quark scale factors    & \NA & 5--14\% & \NA  \\
    Jet energy scale     & \NA & \NA & shape   \\
    Jet energy resolution & \NA & \NA & shape   \\
    Jet mass scale       & \NA & \NA  & 1\% \\
    Jet mass resolution  & \NA & \NA  & 8\% \\
    \PV tagging            & \NA & \NA & 6\% (HP)--11\% (LP) \\
    \PV tagging extrapol.     & \NA & \NA & 8--18\% (HP), 2--8\% (LP) \\
    \cPqb\ tagging            & \NA &  \NA &   3--7\% (1b), 4--5\% (2b)\\
    \cPqb-tagged jet veto    & \NA &  3\% &  1\% \\
    Trigger              & \NA & \NA & 2\%\\
    Lepton identification, isolation      & \NA & \NA & 2\%\\
    $\tau$ lepton identification            & \NA & \NA & 6--8\% ($\ell\tauh$), 10--13\% ($\tauh\tauh$) \\
    $\tau$ lepton identification \pt extrapol.    & \NA & \NA & 0.5--18\% ($\ell\tauh$), 0.2--30\% ($\tauh\tauh$), shape \\
    $\tau$ lepton energy scale & \NA & \NA & 1\% ($\ell\tauh$), 3--5\% ($\tauh\tauh$), shape \\
    Pileup              & \NA & \NA & 0.5\% \\
    Renorm./fact. scales$\dagger$    & \NA & \NA & 2.5--12.5\%($\cPq\cPq$'), 10--19\%($\Pg\Pg$)\\
    PDF yield$\dagger$     & \NA & \NA & 6--37\%($\cPq\cPq$'), 10--64\%($\Pg\Pg$)\\
    PDF acceptance       & \NA & \NA & 0.5--2\% \\
    Integrated luminosity           & \NA & \NA & 2.5\%  \\

  \end{tabular}
\end{table}
\clearpage

\section{Results}
\label{sec:results}

\begin{figure}[!htb]
  \begin{center}
 \includegraphics[width=.495\textwidth]{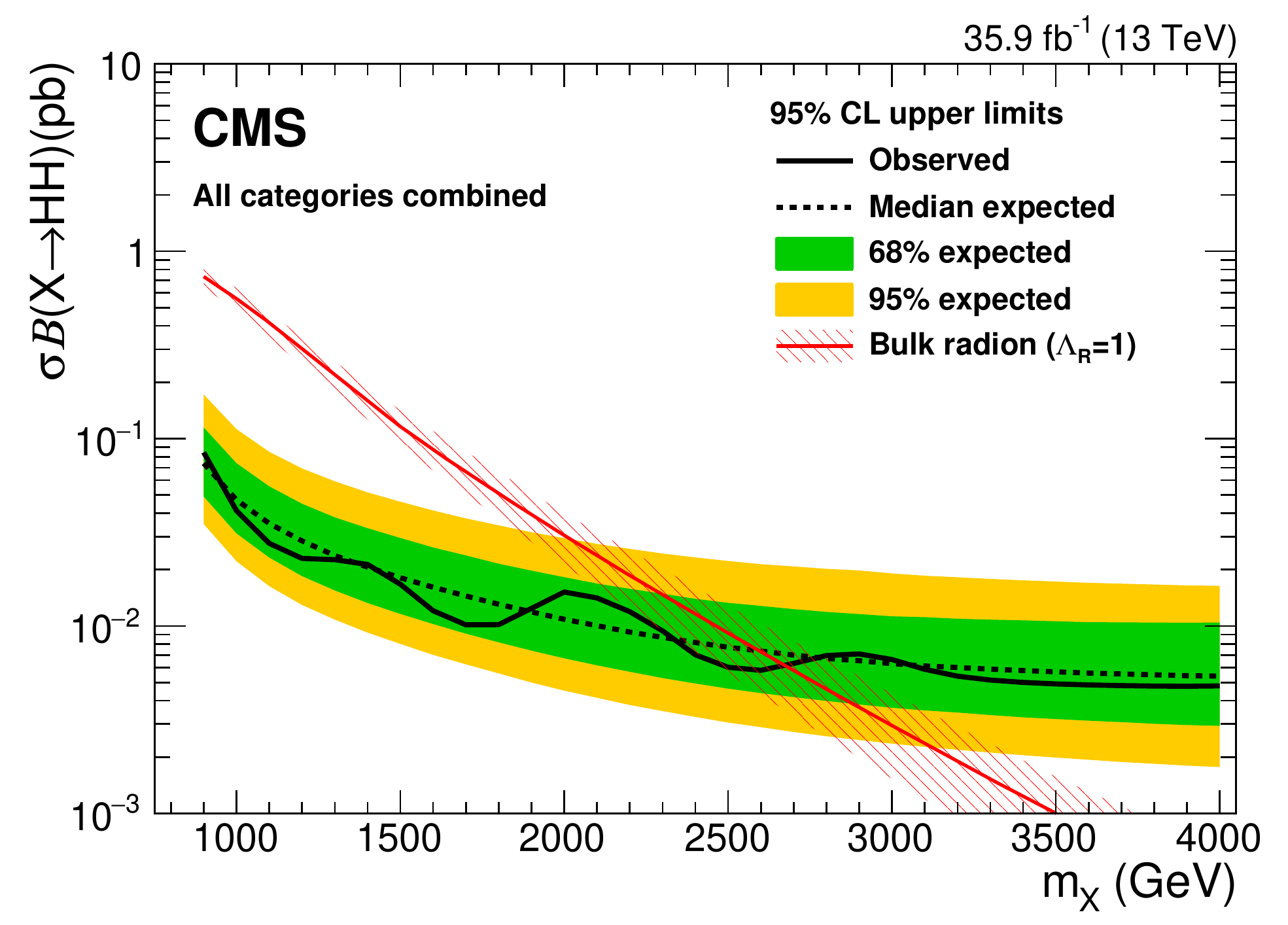}
 \includegraphics[width=.495\textwidth]{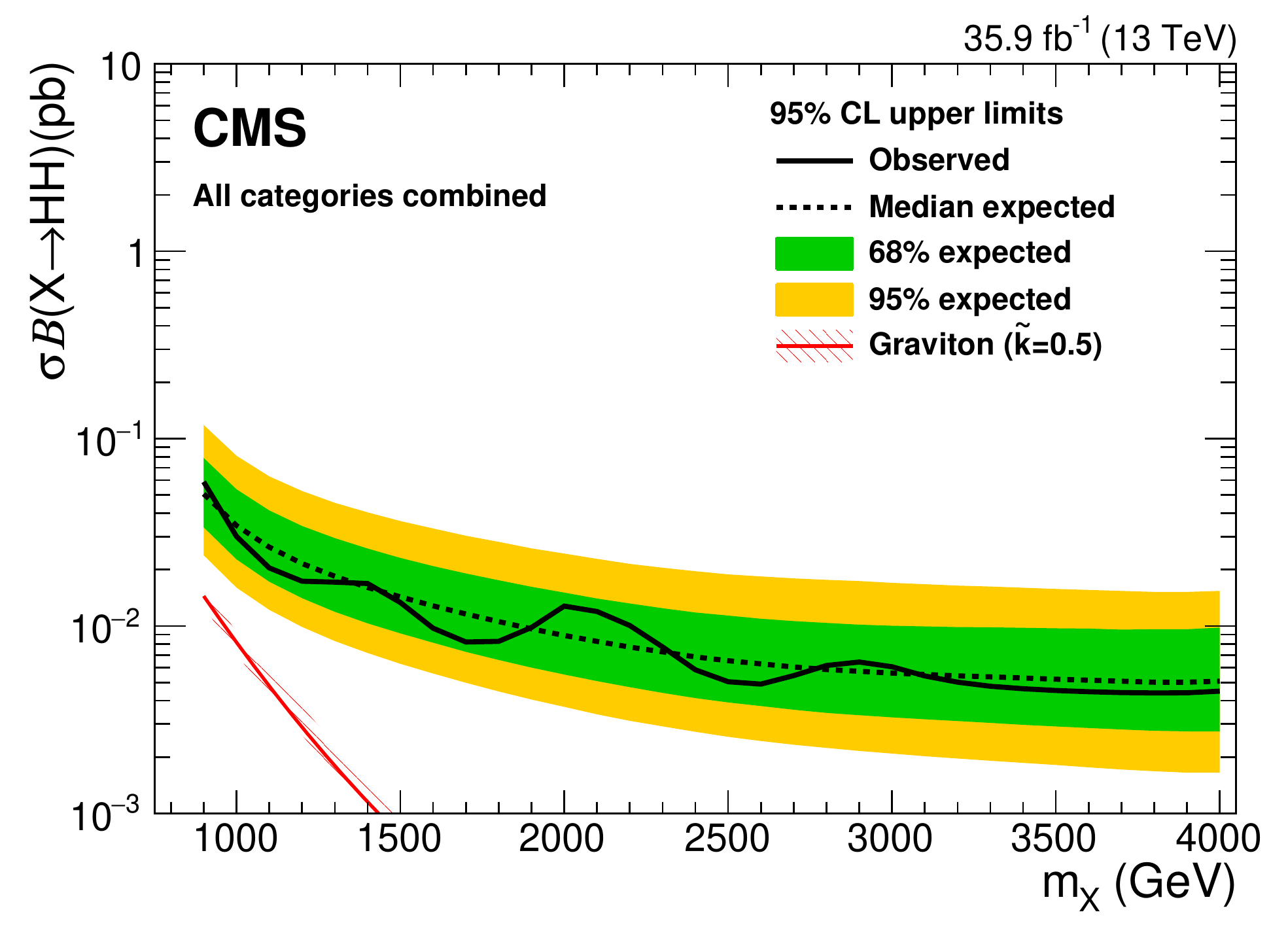}
  \end{center}
  \caption{Observed 95\% \CL upper limits on $\sigma\mathcal{B}$(X(spin-0) $\to\PH\PH$) (\cmsLeft) and $\sigma\mathcal{B}$(X(spin-2) $\to\PH\PH$) (\cmsRight). Expected limits are shown with $\pm$1 and $\pm$2 standard deviation uncertainty bands. The $\ell\tauh$ and $\tauh\tauh$ final states, and the one and two \cPqb-tagged subjet categories are combined, to obtain the limits. The solid red lines and the red dashed areas correspond to the cross sections predicted by the bulk radion and graviton and their corresponding uncertainties, as reported in Table~\ref{tab:Sys}.
}
  \label{fig:Limits_1}
\end{figure}

\begin{figure}[!htb]
  \begin{center}
 \includegraphics[width=.495\textwidth]{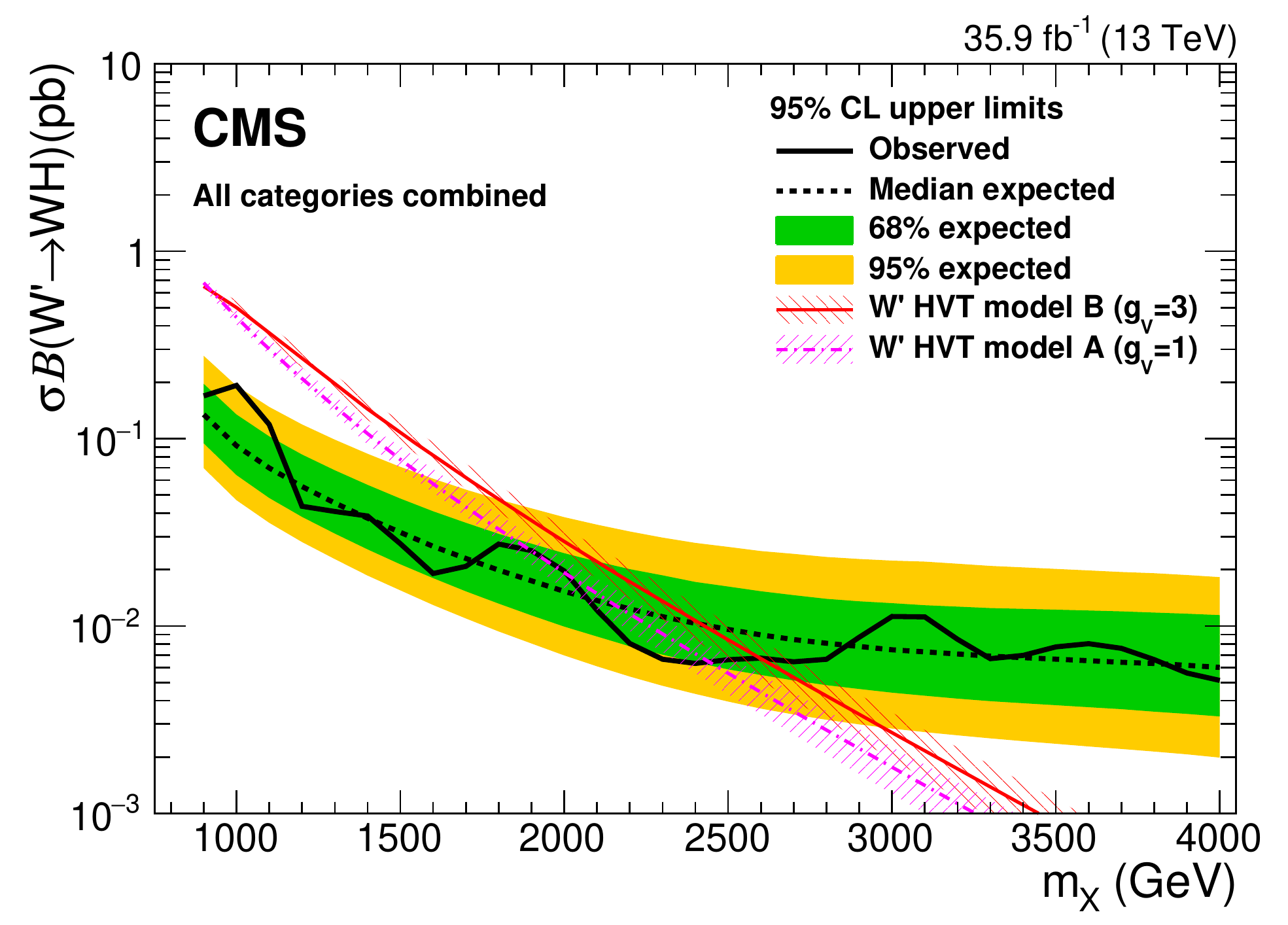}
 \includegraphics[width=.495\textwidth]{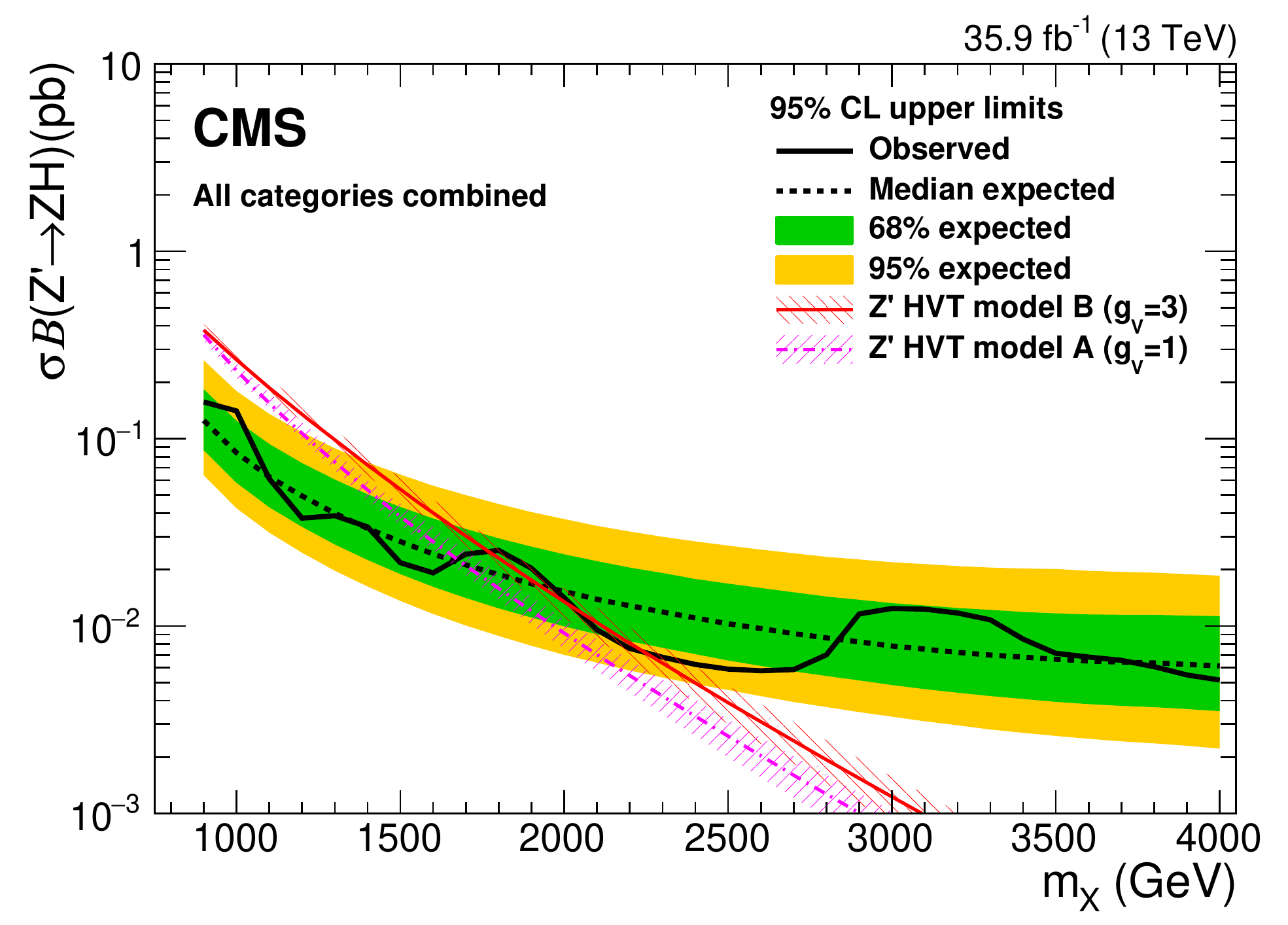}

  \end{center}
  \caption{Observed 95\% \CL upper limits on $\sigma\mathcal{B}$($\PWpr\to\PW\PH$) (\cmsLeft) and $\sigma\mathcal{B}$($\PZpr\to\PZ\PH$) (\cmsRight). Expected limits are shown with $\pm$1 and $\pm$2 standard deviation uncertainty bands. The $\ell\tauh$ and $\tauh\tauh$ final states, for the HP and LP $\tau_{21}$ categories, and the {\PW} and \PZ boson mass signal regions are combined, to obtain the limits. The solid lines and the relative dashed areas in magenta and red correspond to the cross sections predicted by the HVT models A and B, respectively, and their corresponding uncertainties, as reported in Table~\ref{tab:Sys}.
}
  \label{fig:Limits_2}
\end{figure}

Results are obtained from a combined fit of the signal and background to the resonance mass distribution in data, based on a profile likelihood, where the systematic uncertainties are considered as ``nuisance" parameters~\cite{Read:2002hq,AsymptCLs}.
The background-only hypothesis is tested against the signal hypothesis simultaneously in the different categories. No evidence of significant deviations from the background expectation is found.
Assuming that the signals have widths that are negligible
relative to the resonance-mass resolution of approximately 7\%, the 95\% confidence level (\CL) upper limits are determined for the signals using the asymptotic frequentist method~\cite{Junk:1999kv,Read:2002hq,CMS-NOTE-2011-005}. Limits are obtained on the product of the cross section and branching fraction for a heavy resonance (X) that decays to $\PH\PH$, $\PW\PH$, or $\PZ\PH$, as reported in Figs.~\ref{fig:Limits_1}--\ref{fig:Limits_2}.

Resonance spins of 0 and 2 are considered for the $\PH\PH$ final state, while the resonance spin is assumed to be 1 for the $\PW\PH$ and $\PZ\PH$ final states.
For the $\PW\PH$ and $\PZ\PH$ final states, the {\PW} and \PZ boson mass regions are combined because there are contributions from both signals to the two mass regions.
Because of its higher signal selection efficiency, typical sensitivities are better for the $\ell\tauh$ channel than for the $\tauh\tauh$ channel, by a factor that varies between 5 and 2 for resonance masses between 0.9 and 4.0\TeV.
For spin-1 resonances, the HP category has a factor of 4 and 2 better sensitivity than the LP category, for low and high resonance masses, respectively.
For resonances of spin 0 and 2, the 1 b-tagged subjet category has a sensitivity lower by a factor 6 for low resonance masses, compared to the category with 2 b-tagged subjets, but their sensitivities are equal for resonance masses of 4\TeV.
The exclusion limit ranges from 80 to 5\unit{fb} for resonances of spin 0 and 2, and from 180 to 5\unit{fb} for spin-1 resonances.

\begin{figure}[!htb]
  \begin{center}
 \includegraphics[width=.495\textwidth]{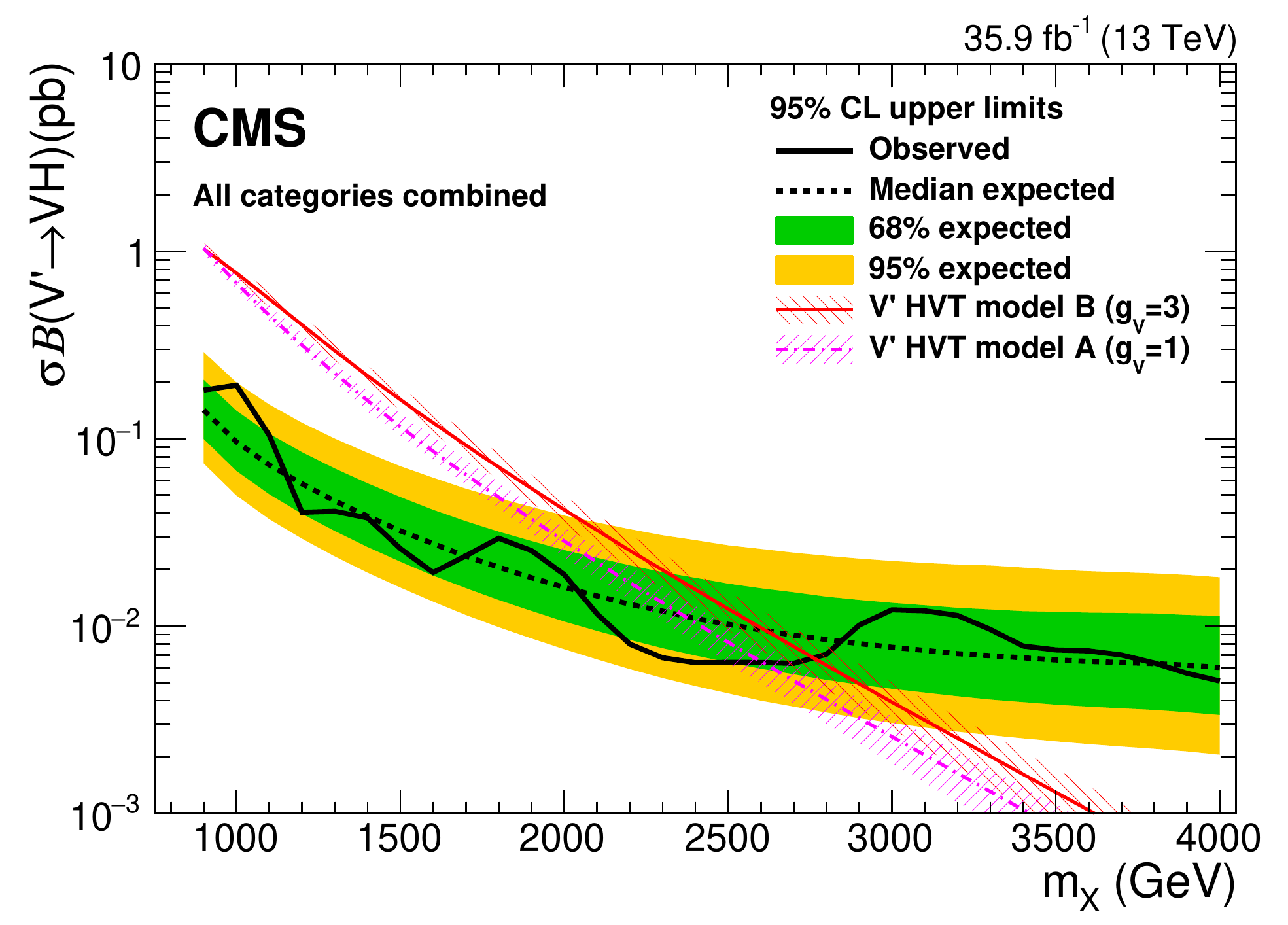}
 \includegraphics[width=.495\textwidth]{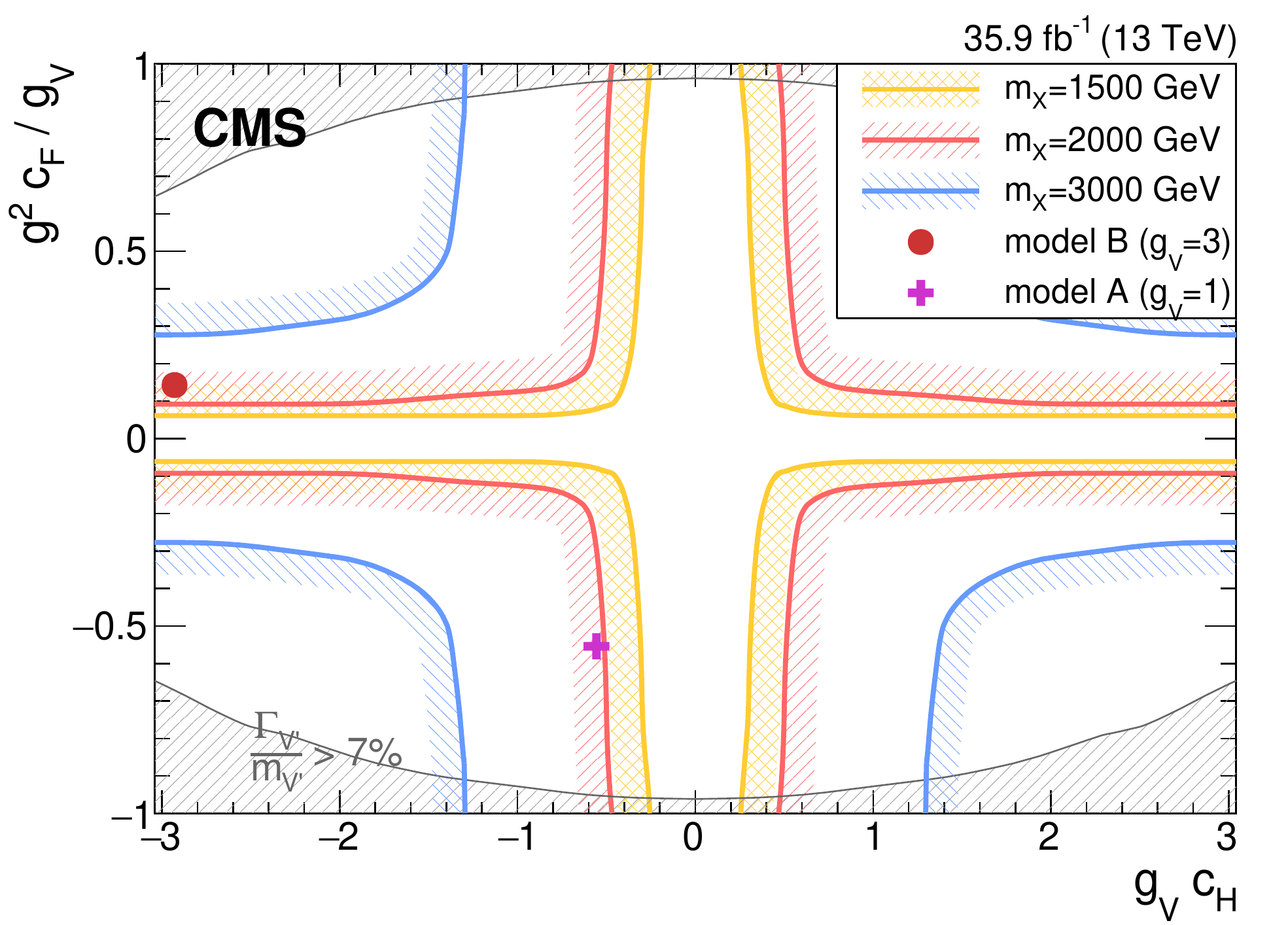}
  \end{center}
  \caption{Expected and observed 95\% \CL upper limit on $\sigma\mathcal{B}$($\PVpr\to\PV\PH$) with $\pm$1 and $\pm$2 standard deviation uncertainty bands (\cmsLeft) in the $\ell\tauh$ and $\tauh\tauh$, $\tau_{21}$ HP and LP categories, with the {\PW} and \PZ boson mass signal regions combined. Observed exclusion limit (\cmsRight) in the space of the HVT model parameters $[ g_{\PV} c_{\PH}, g^2 c_{\mathrm{F}}/g_{\PV}]$, described in the text, for three different mass hypotheses of 1.5, 2.0, and 3.0\TeV. The region of parameter space where the natural resonance width is larger than the typical experimental resolution of 7\%, for which the narrow width assumption is not valid, is shaded in grey.
}
  \label{fig:Limits_3}
\end{figure}

The predictions from bulk radion and graviton models are superimposed on the exclusion limits in Fig.~\ref{fig:Limits_1}, assuming $\Lambda_{\mathrm{R}}=1\TeV$ and $\tilde{k}=0.5$. With this assumption for the theory parameters, a radion resonance with mass below 2.7\TeV is excluded at 95\% \CL.
For a spin-1 signal, the results are interpreted in the context of the simplified HVT benchmark models A and B.
As shown in Fig. \ref{fig:Limits_2}, a \PWpr (\PZpr) resonance of mass lower than 2.6 (1.8)\TeV is excluded at 95\% \CL in the HVT benchmark model B.
The HVT benchmark model A is also reported for completeness. In the mass-degenerate spin-1 triplet hypothesis, the expected and observed limits on the \PVpr resonance are shown in Fig.~\ref{fig:Limits_3} (\cmsLeft).

The exclusion limit shown in Fig.~\ref{fig:Limits_3} (\cmsLeft) can be interpreted as a limit in the space of the HVT model parameters $[ g_{\PV} c_{\PH}, g^2 c_{\mathrm{F}}/g_{\PV}]$. Combining all channels, the excluded region in such a parameter space for narrow resonances is shown in Fig.~\ref{fig:Limits_3} (\cmsRight). The region of parameter space where the natural resonance width is larger than the typical experimental resolution of 7\%, for which the narrow width assumption is not valid, is shaded.

\section{Summary}
\label{sec:conclusions}

A search has been conducted for heavy resonances that decay to two bosons, one of which is a {\PW}, \PZ, or Higgs boson that decays to a pair of quarks, and the other is a Higgs boson that decays to a pair of $\tau$ leptons. The analyzed data are collected by the CMS experiment at $\sqrt{s}=13\TeV$, corresponding to an integrated luminosity of 35.9\fbinv. Reconstruction techniques have been developed to select events in which the $\tau$ lepton pair is highly boosted. The data are consistent with the standard model expectations and upper limits at 95\% confidence level are set on the product of cross section and branching fraction for resonance masses between 0.9 and 4.0\TeV. This search yields the first results at $\sqrt{s}=13\TeV$ for \TeV-scale resonances in the considered mass range and final states.
Assuming the ultraviolet cutoff of the theory $\Lambda_{\mathrm{R}}=1$, Kaluza--Klein excitations of spin-0 radions with mass smaller than 2.7\TeV are excluded at 95\% confidence level.
In the heavy vector triplet model B context, a mass-degenerate vector triplet \PVpr resonance with mass below 2.8\TeV is excluded at 95\% confidence level.

\begin{acknowledgments}

We congratulate our colleagues in the CERN accelerator departments for the excellent performance of the LHC and thank the technical and administrative staffs at CERN and at other CMS institutes for their contributions to the success of the CMS effort. In addition, we gratefully acknowledge the computing centres and personnel of the Worldwide LHC Computing Grid for delivering so effectively the computing infrastructure essential to our analyses. Finally, we acknowledge the enduring support for the construction and operation of the LHC and the CMS detector provided by the following funding agencies: BMWFW and FWF (Austria); FNRS and FWO (Belgium); CNPq, CAPES, FAPERJ, and FAPESP (Brazil); MES (Bulgaria); CERN; CAS, MoST, and NSFC (China); COLCIENCIAS (Colombia); MSES and CSF (Croatia); RPF (Cyprus); SENESCYT (Ecuador); MoER, ERC IUT, and ERDF (Estonia); Academy of Finland, MEC, and HIP (Finland); CEA and CNRS/IN2P3 (France); BMBF, DFG, and HGF (Germany); GSRT (Greece); NKFIA (Hungary); DAE and DST (India); IPM (Iran); SFI (Ireland); INFN (Italy); MSIP and NRF (Republic of Korea); LAS (Lithuania); MOE and UM (Malaysia); BUAP, CINVESTAV, CONACYT, LNS, SEP, and UASLP-FAI (Mexico); MBIE (New Zealand); PAEC (Pakistan); MSHE and NSC (Poland); FCT (Portugal); JINR (Dubna); MON, RosAtom, RAS and RFBR (Russia); MESTD (Serbia); SEIDI, CPAN, PCTI and FEDER (Spain); Swiss Funding Agencies (Switzerland); MST (Taipei); ThEPCenter, IPST, STAR, and NSTDA (Thailand); TUBITAK and TAEK (Turkey); NASU and SFFR (Ukraine); STFC (United Kingdom); DOE and NSF (USA).

\hyphenation{Rachada-pisek} Individuals have received support from the Marie-Curie programme and the European Research Council and Horizon 2020 Grant, contract No. 675440 (European Union); the Leventis Foundation; the A. P. Sloan Foundation; the Alexander von Humboldt Foundation; the Belgian Federal Science Policy Office; the Fonds pour la Formation \`a la Recherche dans l'Industrie et dans l'Agriculture (FRIA-Belgium); the Agentschap voor Innovatie door Wetenschap en Technologie (IWT-Belgium); the F.R.S.-FNRS and FWO (Belgium) under the ``Excellence of Science - EOS" - be.h project n. 30820817; the Ministry of Education, Youth and Sports (MEYS) of the Czech Republic; the Lend\"ulet (``Momentum") Programme and the J\'anos Bolyai Research Scholarship of the Hungarian Academy of Sciences, the New National Excellence Program \'UNKP, the NKFIA research grants 123842, 123959, 124845, 124850 and 125105 (Hungary); the Council of Science and Industrial Research, India; the HOMING PLUS programme of the Foundation for Polish Science, cofinanced from European Union, Regional Development Fund, the Mobility Plus programme of the Ministry of Science and Higher Education, the National Science Center (Poland), contracts Harmonia 2014/14/M/ST2/00428, Opus 2014/13/B/ST2/02543, 2014/15/B/ST2/03998, and 2015/19/B/ST2/02861, Sonata-bis 2012/07/E/ST2/01406; the National Priorities Research Program by Qatar National Research Fund; the Programa Estatal de Fomento de la Investigaci{\'o}n Cient{\'i}fica y T{\'e}cnica de Excelencia Mar\'{\i}a de Maeztu, grant MDM-2015-0509 and the Programa Severo Ochoa del Principado de Asturias; the Thalis and Aristeia programmes cofinanced by EU-ESF and the Greek NSRF; the Rachadapisek Sompot Fund for Postdoctoral Fellowship, Chulalongkorn University and the Chulalongkorn Academic into Its 2nd Century Project Advancement Project (Thailand); the Welch Foundation, contract C-1845; and the Weston Havens Foundation (USA).

\end{acknowledgments}

\bibliography{auto_generated}

\providecommand{\href}[2]{#2}\begingroup\raggedright\begin{thebibliography}{10}%
\makeatletter
\providecommand{\hrefCMSnoop }[0]{\@secondoftwo}%
\makeatother
\providecommand{\doi}{\texttt{doi:}\begingroup \urlstyle{tt}\Url}

\bibitem{Goldberger:1999uk}
\hrefCMSnoop {}{W.~D. Goldberger and M.~B. Wise, ``{Modulus stabilization with
  bulk fields}'',} \textit{ Phys. Rev. Lett.} \textbf{ 83} (1999) 4922,
  \href{http://dx.doi.org/10.1103/PhysRevLett.83.4922}{\doi{10.1103/PhysRevLett.83.4922}},
\href{http://www.arXiv.org/abs/hep-ph/9907447}{\texttt{arXiv:hep-ph/9907447}}.
%%CITATION = HEP-PH/9907447;%%.

\bibitem{DeWolfe:1999cp}
\hrefCMSnoop {}{O.~DeWolfe, D.~Z. Freedman, S.~S. Gubser, and A.~Karch,
  ``{Modeling the fifth-dimension with scalars and gravity}'',} \textit{ Phys.
  Rev. D} \textbf{ 62} (2000) 046008,
  \href{http://dx.doi.org/10.1103/PhysRevD.62.046008}{\doi{10.1103/PhysRevD.62.046008}},
\href{http://www.arXiv.org/abs/hep-th/9909134}{\texttt{arXiv:hep-th/9909134}}.
%%CITATION = HEP-TH/9909134;%%.

\bibitem{Csaki:1999mp}
\hrefCMSnoop {}{C.~Cs{\'a}ki, M.~Graesser, L.~Randall, and J.~Terning,
  ``{Cosmology of brane models with radion stabilization}'',} \textit{ Phys.
  Rev. D} \textbf{ 62} (2000) 045015,
  \href{http://dx.doi.org/10.1103/PhysRevD.62.045015}{\doi{10.1103/PhysRevD.62.045015}},
\href{http://www.arXiv.org/abs/hep-ph/9911406}{\texttt{arXiv:hep-ph/9911406}}.
%%CITATION = HEP-PH/9911406;%%.

\bibitem{Randall:1999ee}
\hrefCMSnoop {}{L.~Randall and R.~Sundrum, ``{A large mass hierarchy from a
  small extra dimension}'',} \textit{ Phys. Rev. Lett.} \textbf{ 83} (1999)
  3370,
  \href{http://dx.doi.org/10.1103/PhysRevLett.83.3370}{\doi{10.1103/PhysRevLett.83.3370}},
\href{http://www.arXiv.org/abs/hep-ph/9905221}{\texttt{arXiv:hep-ph/9905221}}.
%%CITATION = HEP-PH/9905221;%%.

\bibitem{Randall:1999vf}
\hrefCMSnoop {}{L.~Randall and R.~Sundrum, ``{An alternative to
  compactification}'',} \textit{ Phys. Rev. Lett.} \textbf{ 83} (1999) 4690,
  \href{http://dx.doi.org/10.1103/PhysRevLett.83.4690}{\doi{10.1103/PhysRevLett.83.4690}},
\href{http://www.arXiv.org/abs/hep-th/9906064}{\texttt{arXiv:hep-th/9906064}}.
%%CITATION = HEP-TH/9906064;%%.

\bibitem{Agashe:2007zd}
\hrefCMSnoop {}{K.~Agashe, H.~Davoudiasl, G.~Perez, and A.~Soni, ``{Warped
  gravitons at the LHC and beyond}'',} \textit{ Phys. Rev. D} \textbf{ 76}
  (2007) 036006,
  \href{http://dx.doi.org/10.1103/PhysRevD.76.036006}{\doi{10.1103/PhysRevD.76.036006}},
\href{http://www.arXiv.org/abs/hep-ph/0701186}{\texttt{arXiv:hep-ph/0701186}}.
%%CITATION = HEP-PH/0701186;%%.

\bibitem{Fitzpatrick:2007qr}
\hrefCMSnoop {}{A.~L. Fitzpatrick, J.~Kaplan, L.~Randall, and L.-T. Wang,
  ``{Searching for the Kaluza-Klein graviton in bulk RS Models}'',} \textit{
  JHEP} \textbf{ 09} (2007) 013,
  \href{http://dx.doi.org/10.1088/1126-6708/2007/09/013}{\doi{10.1088/1126-6708/2007/09/013}},
\href{http://www.arXiv.org/abs/hep-ph/0701150}{\texttt{arXiv:hep-ph/0701150}}.
%%CITATION = HEP-PH/0701150;%%.

\bibitem{Antipin:2007pi}
\hrefCMSnoop {}{O.~Antipin, D.~Atwood, and A.~Soni, ``{Search for RS gravitons
  via W$_L$ W$_L$ decays}'',} \textit{ Phys. Lett. B} \textbf{ 666} (2008) 155,
  \href{http://dx.doi.org/10.1016/j.physletb.2008.07.009}{\doi{10.1016/j.physletb.2008.07.009}},
\href{http://www.arXiv.org/abs/0711.3175}{\texttt{arXiv:0711.3175}}.
%%CITATION = ARXIV:0711.3175;%%.

\bibitem{Bellazzini:2014yua}
\hrefCMSnoop {}{B.~Bellazzini, C.~Cs{\'a}ki, and J.~Serra, ``{Composite
  Higgses}'',} \textit{ Eur. Phys. J. C} \textbf{ 74} (2014) 2766,
  \href{http://dx.doi.org/10.1140/epjc/s10052-014-2766-x}{\doi{10.1140/epjc/s10052-014-2766-x}},
\href{http://www.arXiv.org/abs/1401.2457}{\texttt{arXiv:1401.2457}}.
%%CITATION = ARXIV:1401.2457;%%.

\bibitem{CHM2}
\hrefCMSnoop {}{R.~Contino, D.~Marzocca, D.~Pappadopulo, and R.~Rattazzi, ``{On
  the effect of resonances in composite Higgs phenomenology}'',} \textit{ JHEP}
  \textbf{ 10} (2011) 081,
  \href{http://dx.doi.org/10.1007/JHEP10(2011)081}{\doi{10.1007/JHEP10(2011)081}},
\href{http://www.arXiv.org/abs/1109.1570}{\texttt{arXiv:1109.1570}}.
%%CITATION = ARXIV:1109.1570;%%.

\bibitem{Composite2}
\hrefCMSnoop {}{D.~Marzocca, M.~Serone, and J.~Shu, ``General composite {Higgs}
  models'',} \textit{ JHEP} \textbf{ 08} (2012) 013,
  \href{http://dx.doi.org/10.1007/JHEP08(2012)013}{\doi{10.1007/JHEP08(2012)013}},
\href{http://www.arXiv.org/abs/1205.0770}{\texttt{arXiv:1205.0770}}.
%%CITATION = ARXIV:1205.0770;%%.

\bibitem{Greco:2014aza}
\hrefCMSnoop {}{D.~Greco and D.~Liu, ``{Hunting composite vector resonances at
  the LHC: naturalness facing data}'',} \textit{ JHEP} \textbf{ 12} (2014) 126,
  \href{http://dx.doi.org/10.1007/JHEP12(2014)126}{\doi{10.1007/JHEP12(2014)126}},
\href{http://www.arXiv.org/abs/1410.2883}{\texttt{arXiv:1410.2883}}.
%%CITATION = ARXIV:1410.2883;%%.

\bibitem{Schmaltz:2005ky}
\hrefCMSnoop {}{M.~Schmaltz and D.~Tucker-Smith, ``{Little Higgs review}'',}
  \textit{ Ann. Rev. Nucl. Part. Sci.} \textbf{ 55} (2005) 229,
  \href{http://dx.doi.org/10.1146/annurev.nucl.55.090704.151502}{\doi{10.1146/annurev.nucl.55.090704.151502}},
\href{http://www.arXiv.org/abs/hep-ph/0502182}{\texttt{arXiv:hep-ph/0502182}}.
%%CITATION = HEP-PH/0502182;%%.

\bibitem{ArkaniHamed:2002qy}
\hrefCMSnoop {}{N.~Arkani-Hamed, A.~G. Cohen, E.~Katz, and A.~E. Nelson, ``The
  littlest {Higgs}'',} \textit{ JHEP} \textbf{ 07} (2002) 034,
  \href{http://dx.doi.org/10.1088/1126-6708/2002/07/034}{\doi{10.1088/1126-6708/2002/07/034}},
\href{http://www.arXiv.org/abs/hep-ph/0206021}{\texttt{arXiv:hep-ph/0206021}}.
%%CITATION = HEP-PH/0206021;%%.

\bibitem{Altarelli}
\hrefCMSnoop {}{G.~Altarelli, B.~Mele, and M.~Ruiz-Altaba, ``Searching for new
  heavy vector bosons in $\rm p\bar{p}$ colliders'',} \textit{ Z. Phys. C}
  \textbf{ 45} (1989) 109,
  \href{http://dx.doi.org/10.1007/BF01556677}{\doi{10.1007/BF01556677}}.

\bibitem{Pappadopulo:2014qza}
\hrefCMSnoop {}{D.~Pappadopulo, A.~Thamm, R.~Torre, and A.~Wulzer, ``{Heavy
  vector triplets: bridging theory and data}'',} \textit{ JHEP} \textbf{ 09}
  (2014) 060,
  \href{http://dx.doi.org/10.1007/JHEP09(2014)060}{\doi{10.1007/JHEP09(2014)060}},
\href{http://www.arXiv.org/abs/1402.4431}{\texttt{arXiv:1402.4431}}.
%%CITATION = ARXIV:1402.4431;%%.

\bibitem{Gouzevitch:2013qca}
M.~Gouzevitch\hrefCMSnoop {}{ {et~al.}, ``{Scale-invariant resonance tagging in
  multijet events and new physics in Higgs pair production}'',} \textit{ JHEP}
  \textbf{ 07} (2013) 148,
  \href{http://dx.doi.org/10.1007/JHEP07(2013)148}{\doi{10.1007/JHEP07(2013)148}},
\href{http://www.arXiv.org/abs/1303.6636}{\texttt{arXiv:1303.6636}}.
%%CITATION = ARXIV:1303.6636;%%.

\bibitem{Aaboud:2017cxo}
\hrefCMSnoop {}{{ATLAS Collaboration}, ``{Search for heavy resonances decaying
  into a W or Z boson and a Higgs boson in final states with leptons and b-jets
  in 36 fb$^{-1}$ of $\sqrt{s} = 13 \TeV$ pp collisions with the ATLAS
  detector}'',} \textit{ JHEP} \textbf{ 03} (2018) 174,
  \href{http://dx.doi.org/10.1007/JHEP03(2018)174}{\doi{10.1007/JHEP03(2018)174}},
\href{http://www.arXiv.org/abs/1712.06518}{\texttt{arXiv:1712.06518}}.
%%CITATION = ARXIV:1712.06518;%%.

\bibitem{Sirunyan:2018qob}
\hrefCMSnoop {}{{CMS Collaboration}, ``{Search for heavy resonances decaying
  into a vector boson and a Higgs boson in final states with charged leptons,
  neutrinos and b quarks at $\sqrt{s}=13 \TeV$}'',} (2018).
  \href{http://www.arXiv.org/abs/1807.02826}{\texttt{arXiv:1807.02826}}.
Submitted to {JHEP}.
%%CITATION = ARXIV:1807.02826;%%.

\bibitem{Aaboud:2017ahz}
\hrefCMSnoop {}{{ATLAS Collaboration}, ``{Search for heavy resonances decaying
  to a W or Z boson and a Higgs boson in the
  $\mathrm{q\bar{q}^{(\prime)}b\bar{b}}$ final state in pp collisions at
  $\sqrt{s} = 13$ TeV with the ATLAS detector}'',} \textit{ Phys. Lett. B}
  \textbf{ 774} (2017) 494,
  \href{http://dx.doi.org/10.1016/j.physletb.2017.09.066}{\doi{10.1016/j.physletb.2017.09.066}},
\href{http://www.arXiv.org/abs/1707.06958}{\texttt{arXiv:1707.06958}}.
%%CITATION = ARXIV:1707.06958;%%.

\bibitem{Sirunyan:2018ivv}
\hrefCMSnoop {}{{CMS Collaboration}, ``{Search for a heavy resonance decaying
  into a Z boson and a vector boson in the $ \nu
  \overline{\nu}\mathrm{q}\overline{\mathrm{q}} $ final state}'',} \textit{
  JHEP} \textbf{ 07} (2018) 075,
  \href{http://dx.doi.org/10.1007/JHEP07(2018)075}{\doi{10.1007/JHEP07(2018)075}},
\href{http://www.arXiv.org/abs/1803.03838}{\texttt{arXiv:1803.03838}}.
%%CITATION = ARXIV:1803.03838;%%.

\bibitem{Sirunyan:2018hsl}
\hrefCMSnoop {}{{CMS Collaboration}, ``{Search for a heavy resonance decaying
  into a Z boson and a Z or W boson in $2\ell2$q final states at $\sqrt{s}=13
  \TeV$}'',} \textit{ JHEP} \textbf{ 09} (2018) 101,
  \href{http://dx.doi.org/10.1007/JHEP09(2018)101}{\doi{10.1007/JHEP09(2018)101}},
\href{http://www.arXiv.org/abs/1803.10093}{\texttt{arXiv:1803.10093}}.
%%CITATION = ARXIV:1803.10093;%%.

\bibitem{PhysRevD.97.072006}
\hrefCMSnoop {}{{CMS Collaboration}, ``{Search for massive resonances decaying
  into WW, WZ, ZZ, qW, and qZ with dijet final states at $\sqrt{s}=13\TeV$}'',}
  \textit{ Phys. Rev. D} \textbf{ 97} (Apr, 2018) 072006,
  \href{http://dx.doi.org/10.1103/PhysRevD.97.072006}{\doi{10.1103/PhysRevD.97.072006}}.

\bibitem{Aaboud:2017itg}
\hrefCMSnoop {}{{ATLAS Collaboration}, ``{Searches for heavy ZZ and ZW
  resonances in the $\ell\ell$ qq and $\nu\nu$ qq final states in pp collisions
  at $\sqrt{s}=13$ TeV with the ATLAS detector}'',} \textit{ JHEP} \textbf{ 03}
  (2018) 009,
  \href{http://dx.doi.org/10.1007/JHEP03(2018)009}{\doi{10.1007/JHEP03(2018)009}},
\href{http://www.arXiv.org/abs/1708.09638}{\texttt{arXiv:1708.09638}}.
%%CITATION = ARXIV:1708.09638;%%.

\bibitem{Aaboud:2017eta}
\hrefCMSnoop {}{{ATLAS Collaboration}, ``{Search for diboson resonances with
  boson-tagged jets in pp collisions at $\sqrt{s}=13$ TeV with the ATLAS
  detector}'',} \textit{ Phys. Lett. B} \textbf{ 777} (2018) 91,
  \href{http://dx.doi.org/10.1016/j.physletb.2017.12.011}{\doi{10.1016/j.physletb.2017.12.011}},
\href{http://www.arXiv.org/abs/1708.04445}{\texttt{arXiv:1708.04445}}.
%%CITATION = ARXIV:1708.04445;%%.

\bibitem{Aaboud:2017fgj}
\hrefCMSnoop {}{{ATLAS Collaboration}, ``{Search for WW/WZ resonance production
  in $\ell \nu$ qq final states in pp collisions at $\sqrt{s} =$ 13 TeV with
  the ATLAS detector}'',} \textit{ JHEP} \textbf{ 03} (2018) 042,
  \href{http://dx.doi.org/10.1007/JHEP03(2018)042}{\doi{10.1007/JHEP03(2018)042}},
\href{http://www.arXiv.org/abs/1710.07235}{\texttt{arXiv:1710.07235}}.
%%CITATION = ARXIV:1710.07235;%%.

\bibitem{Aaboud:2017gsl}
\hrefCMSnoop {}{{ATLAS Collaboration}, ``{Search for heavy resonances decaying
  into WW in the $e\nu\mu\nu$ final state in pp collisions at $\sqrt{s}=13$ TeV
  with the ATLAS detector}'',} \textit{ Eur. Phys. J. C} \textbf{ 78} (2018),
  no.~1, 24,
  \href{http://dx.doi.org/10.1140/epjc/s10052-017-5491-4}{\doi{10.1140/epjc/s10052-017-5491-4}},
\href{http://www.arXiv.org/abs/1710.01123}{\texttt{arXiv:1710.01123}}.
%%CITATION = ARXIV:1710.01123;%%.

\bibitem{Aaboud:2018ohp}
\hrefCMSnoop {}{{ATLAS Collaboration}, ``{Search for resonant WZ production in
  the fully leptonic final state in proton-proton collisions at $\sqrt{s} = 13$
  TeV with the ATLAS detector}'',} \textit{ Phys. Lett. B} \textbf{ 787} (2018)
  68,
  \href{http://dx.doi.org/10.1016/j.physletb.2018.10.021}{\doi{10.1016/j.physletb.2018.10.021}},
\href{http://www.arXiv.org/abs/1806.01532}{\texttt{arXiv:1806.01532}}.
%%CITATION = ARXIV:1806.01532;%%.

\bibitem{Sirunyan:2017jtu}
\hrefCMSnoop {}{{CMS Collaboration}, ``{Search for ZZ resonances in the
  2$\ell$2$\nu$ final state in proton-proton collisions at 13 TeV}'',} \textit{
  JHEP} \textbf{ 03} (2018) 003,
  \href{http://dx.doi.org/10.1007/JHEP03(2018)003}{\doi{10.1007/JHEP03(2018)003}},
\href{http://www.arXiv.org/abs/1711.04370}{\texttt{arXiv:1711.04370}}.
%%CITATION = ARXIV:1711.04370;%%.

\bibitem{Sirunyan:2017isc}
\hrefCMSnoop {}{{CMS Collaboration}, ``{Search for a massive resonance decaying
  to a pair of Higgs bosons in the four b quark final state in proton-proton
  collisions at $\sqrt{s}=$ 13 TeV}'',} \textit{ Phys. Lett. B} \textbf{ 781}
  (2018) 244,
  \href{http://dx.doi.org/10.1016/j.physletb.2018.03.084}{\doi{10.1016/j.physletb.2018.03.084}},
\href{http://www.arXiv.org/abs/1710.04960}{\texttt{arXiv:1710.04960}}.
%%CITATION = ARXIV:1710.04960;%%.

\bibitem{Aaboud:2018knk}
\hrefCMSnoop {}{{ATLAS Collaboration}, ``{Search for pair production of Higgs
  bosons in the $b\bar{b}b\bar{b}$ final state using proton-proton collisions
  at $\sqrt{s} = 13$ TeV with the ATLAS detector}'',} (2018).
  \href{http://www.arXiv.org/abs/1804.06174}{\texttt{arXiv:1804.06174}}.
Submitted to {JHEP}.
%%CITATION = ARXIV:1804.06174;%%.

\bibitem{Sirunyan:2017nrt}
\hrefCMSnoop {}{{CMS Collaboration}, ``{Combination of searches for heavy
  resonances decaying to WW, WZ, ZZ, WH, and ZH boson pairs in proton-proton
  collisions at $\sqrt{s}=8$ and $13$ TeV}'',} \textit{ Phys. Lett. B} \textbf{
  774} (2017) 533,
  \href{http://dx.doi.org/10.1016/j.physletb.2017.09.083}{\doi{10.1016/j.physletb.2017.09.083}},
\href{http://www.arXiv.org/abs/1705.09171}{\texttt{arXiv:1705.09171}}.
%%CITATION = ARXIV:1705.09171;%%.

\bibitem{Aaboud:2018bun}
\hrefCMSnoop {}{{ATLAS Collaboration}, ``{Combination of searches for heavy
  resonances decaying into bosonic and leptonic final states using 36 fb$^{-1}$
  of proton-proton collision data at $\sqrt{s} = 13$ TeV with the ATLAS
  detector}'',} \textit{ Phys. Rev. D} \textbf{ 98} (2018) 052008,
  \href{http://dx.doi.org/10.1103/PhysRevD.98.052008}{\doi{10.1103/PhysRevD.98.052008}},
\href{http://www.arXiv.org/abs/1808.02380}{\texttt{arXiv:1808.02380}}.
%%CITATION = ARXIV:1808.02380;%%.

\bibitem{Khachatryan:2015ywa}
\hrefCMSnoop {}{{CMS Collaboration}, ``{Search for narrow high-mass resonances
  in proton-proton collisions at $\sqrt{s}=8\TeV$ decaying to a Z and a Higgs
  boson}'',} \textit{ Phys. Lett. B} \textbf{ 748} (2015) 255,
  \href{http://dx.doi.org/10.1016/j.physletb.2015.07.011}{\doi{10.1016/j.physletb.2015.07.011}},
\href{http://www.arXiv.org/abs/1502.04994}{\texttt{arXiv:1502.04994}}.
%%CITATION = ARXIV:1502.04994;%%.

\bibitem{Sirunyan:2017djm}
\hrefCMSnoop {}{{CMS Collaboration}, ``{Search for Higgs boson pair production
  in events with two bottom quarks and two tau leptons in proton-proton
  collisions at $\sqrt{s}= 13\TeV$}'',} \textit{ Phys. Lett. B} \textbf{ 778}
  (2018) 101,
  \href{http://dx.doi.org/10.1016/j.physletb.2018.01.001}{\doi{10.1016/j.physletb.2018.01.001}},
\href{http://www.arXiv.org/abs/1707.02909}{\texttt{arXiv:1707.02909}}.
%%CITATION = ARXIV:1707.02909;%%.

\bibitem{Aaboud:2018sfw}
\hrefCMSnoop {}{{ATLAS Collaboration}, ``{A search for resonant and
  non-resonant Higgs boson pair production in the
  $\mathrm{b\bar{b}\tau^+\tau^-}$ decay channelin pp collisions at
  $\sqrt{s}=13$ TeV with the ATLAS detector}'',} (2018).
  \href{http://www.arXiv.org/abs/1808.00336}{\texttt{arXiv:1808.00336}}.
Submitted to {Phys. Rev. Lett.}
%%CITATION = ARXIV:1808.00336;%%.

\bibitem{Khachatryan:2016bia}
\hrefCMSnoop {}{{CMS Collaboration}, ``{The CMS trigger system}'',} \textit{
  JINST} \textbf{ 12} (2017) P01020,
  \href{http://dx.doi.org/10.1088/1748-0221/12/01/P01020}{\doi{10.1088/1748-0221/12/01/P01020}},
\href{http://www.arXiv.org/abs/1609.02366}{\texttt{arXiv:1609.02366}}.
%%CITATION = ARXIV:1609.02366;%%.

\bibitem{Chatrchyan:2008aa}
\hrefCMSnoop {}{{CMS Collaboration}, ``{The CMS experiment at the CERN LHC}'',}
  \textit{ JINST} \textbf{ 3} (2008) S08004,
\href{http://dx.doi.org/10.1088/1748-0221/3/08/S08004}{\doi{10.1088/1748-0221/3/08/S08004}}.
%%CITATION = JINST,3,S08004;%%.

\bibitem{Alwall:2014hca}
J.~Alwall\hrefCMSnoop {}{ {et~al.}, ``{The automated computation of tree-level
  and next-to-leading order differential cross sections, and their matching to
  parton shower simulations}'',} \textit{ JHEP} \textbf{ 07} (2014) 079,
  \href{http://dx.doi.org/10.1007/JHEP07(2014)079}{\doi{10.1007/JHEP07(2014)079}},
\href{http://www.arXiv.org/abs/1405.0301}{\texttt{arXiv:1405.0301}}.
%%CITATION = ARXIV:1405.0301;%%.

\bibitem{Nason:2004rx}
\hrefCMSnoop {}{P.~Nason, ``{A new method for combining NLO QCD with shower
  Monte Carlo algorithms}'',} \textit{ JHEP} \textbf{ 11} (2004) 040,
  \href{http://dx.doi.org/10.1088/1126-6708/2004/11/040}{\doi{10.1088/1126-6708/2004/11/040}},
\href{http://www.arXiv.org/abs/hep-ph/0409146}{\texttt{arXiv:hep-ph/0409146}}.
%%CITATION = HEP-PH/0409146;%%.

\bibitem{Frixione:2007vw}
\hrefCMSnoop {}{S.~Frixione, P.~Nason, and C.~Oleari, ``{Matching NLO QCD
  computations with Parton Shower simulations: the POWHEG method}'',} \textit{
  JHEP} \textbf{ 11} (2007) 070,
  \href{http://dx.doi.org/10.1088/1126-6708/2007/11/070}{\doi{10.1088/1126-6708/2007/11/070}},
\href{http://www.arXiv.org/abs/0709.2092}{\texttt{arXiv:0709.2092}}.
%%CITATION = ARXIV:0709.2092;%%.

\bibitem{Alioli:2010xd}
\hrefCMSnoop {}{S.~Alioli, P.~Nason, C.~Oleari, and E.~Re, ``{A general
  framework for implementing NLO calculations in shower Monte Carlo programs:
  the POWHEG BOX}'',} \textit{ JHEP} \textbf{ 06} (2010) 043,
  \href{http://dx.doi.org/10.1007/JHEP06(2010)043}{\doi{10.1007/JHEP06(2010)043}},
\href{http://www.arXiv.org/abs/1002.2581}{\texttt{arXiv:1002.2581}}.
%%CITATION = ARXIV:1002.2581;%%.

\bibitem{Alioli:2011as}
\hrefCMSnoop {}{S.~Alioli, S.-O. Moch, and P.~Uwer, ``{Hadronic top-quark
  pair-production with one jet and parton showering}'',} \textit{ JHEP}
  \textbf{ 01} (2012) 137,
  \href{http://dx.doi.org/10.1007/JHEP01(2012)137}{\doi{10.1007/JHEP01(2012)137}},
\href{http://www.arXiv.org/abs/1110.5251}{\texttt{arXiv:1110.5251}}.
%%CITATION = ARXIV:1110.5251;%%.

\bibitem{Sjostrand:2014zea}
T.~Sj{\"o}strand\hrefCMSnoop {}{ {et~al.}, ``{An Introduction to PYTHIA
  8.2}'',} \textit{ Comput. Phys. Commun.} \textbf{ 191} (2015) 159,
  \href{http://dx.doi.org/10.1016/j.cpc.2015.01.024}{\doi{10.1016/j.cpc.2015.01.024}},
\href{http://www.arXiv.org/abs/1410.3012}{\texttt{arXiv:1410.3012}}.
%%CITATION = ARXIV:1410.3012;%%.

\bibitem{Davidson:2010rw}
N.~Davidson\hrefCMSnoop {}{ {et~al.}, ``{Universal interface of TAUOLA
  technical and physics documentation}'',} \textit{ Comput. Phys. Commun.}
  \textbf{ 183} (2012) 821,
  \href{http://dx.doi.org/10.1016/j.cpc.2011.12.009}{\doi{10.1016/j.cpc.2011.12.009}},
\href{http://www.arXiv.org/abs/1002.0543}{\texttt{arXiv:1002.0543}}.
%%CITATION = ARXIV:1002.0543;%%.

\bibitem{Agostinelli:2002hh}
\hrefCMSnoop {}{{GEANT4} Collaboration, ``{\GEANTfour}---a simulation
  toolkit'',} \textit{ Nucl. Instrum. Meth. A} \textbf{ 506} (2003) 250,
\href{http://dx.doi.org/10.1016/S0168-9002(03)01368-8}{\doi{10.1016/S0168-9002(03)01368-8}}.
%%CITATION = NUIMA,A506,250;%%.

\bibitem{Sirunyan:2017ulk}
\hrefCMSnoop {}{{CMS Collaboration}, ``{Particle-flow reconstruction and global
  event description with the CMS detector}'',} \textit{ JINST} \textbf{ 12}
  (2017) P10003,
  \href{http://dx.doi.org/10.1088/1748-0221/12/10/P10003}{\doi{10.1088/1748-0221/12/10/P10003}},
\href{http://www.arXiv.org/abs/1706.04965}{\texttt{arXiv:1706.04965}}.
%%CITATION = ARXIV:1706.04965;%%.

\bibitem{Cacciari:2008gp}
\hrefCMSnoop {}{M.~Cacciari, G.~P. Salam, and G.~Soyez, ``{The anti-\kt jet
  clustering algorithm}'',} \textit{ JHEP} \textbf{ 04} (2008) 063,
  \href{http://dx.doi.org/10.1088/1126-6708/2008/04/063}{\doi{10.1088/1126-6708/2008/04/063}},
\href{http://www.arXiv.org/abs/0802.1189}{\texttt{arXiv:0802.1189}}.
%%CITATION = ARXIV:0802.1189;%%.

\bibitem{Cacciari:2011ma}
\hrefCMSnoop {}{M.~Cacciari, G.~P. Salam, and G.~Soyez, ``{FastJet user
  manual}'',} \textit{ Eur. Phys. J. C} \textbf{ 72} (2012) 1896,
  \href{http://dx.doi.org/10.1140/epjc/s10052-012-1896-2}{\doi{10.1140/epjc/s10052-012-1896-2}},
\href{http://www.arXiv.org/abs/1111.6097}{\texttt{arXiv:1111.6097}}.
%%CITATION = ARXIV:1111.6097;%%.

\bibitem{Khachatryan:2016kdb}
\hrefCMSnoop {}{{CMS Collaboration}, ``{Jet energy scale and resolution in the
  CMS experiment in pp collisions at 8 TeV}'',} \textit{ JINST} \textbf{ 12}
  (2017) P02014,
  \href{http://dx.doi.org/10.1088/1748-0221/12/02/P02014}{\doi{10.1088/1748-0221/12/02/P02014}},
\href{http://www.arXiv.org/abs/1607.03663}{\texttt{arXiv:1607.03663}}.
%%CITATION = ARXIV:1607.03663;%%.

\bibitem{CMS:2016ljj}
\href {http://cds.cern.ch/record/1479660}{{CMS Collaboration}, ``{Performance
  of missing energy reconstruction in 13 TeV pp collision data using the CMS
  detector}'',} CMS Physics Analysis Summary CMS-PAS-JME-16-004, CERN, 2016.

\bibitem{Bertolini2014}
\hrefCMSnoop {}{D.~Bertolini, P.~Harris, M.~Low, and N.~Tran, ``{Pileup per
  particle identification}'',} \textit{ JHEP} \textbf{ 10} (2014) 059,
  \href{http://dx.doi.org/10.1007/JHEP10(2014)059}{\doi{10.1007/JHEP10(2014)059}},
\href{http://www.arXiv.org/abs/1407.6013}{\texttt{arXiv:1407.6013}}.
%%CITATION = ARXIV:1407.6013;%%.

\bibitem{Dasgupta:2013ihk}
\hrefCMSnoop {}{M.~Dasgupta, A.~Fregoso, S.~Marzani, and G.~P. Salam,
  ``{Towards an understanding of jet substructure}'',} \textit{ JHEP} \textbf{
  09} (2013) 029,
  \href{http://dx.doi.org/10.1007/JHEP09(2013)029}{\doi{10.1007/JHEP09(2013)029}},
\href{http://www.arXiv.org/abs/1307.0007}{\texttt{arXiv:1307.0007}}.
%%CITATION = ARXIV:1307.0007;%%.

\bibitem{Larkoski:2014wba}
\hrefCMSnoop {}{A.~J. Larkoski, S.~Marzani, G.~Soyez, and J.~Thaler, ``Soft
  drop'',} \textit{ JHEP} \textbf{ 05} (2014) 146,
  \href{http://dx.doi.org/10.1007/JHEP05(2014)146}{\doi{10.1007/JHEP05(2014)146}},
\href{http://www.arXiv.org/abs/1402.2657}{\texttt{arXiv:1402.2657}}.
%%CITATION = ARXIV:1402.2657;%%.

\bibitem{Thaler:2010tr}
\hrefCMSnoop {}{J.~Thaler and K.~Van~Tilburg, ``{Identifying boosted objects
  with N-subjettiness}'',} \textit{ JHEP} \textbf{ 03} (2011) 015,
  \href{http://dx.doi.org/10.1007/JHEP03(2011)015}{\doi{10.1007/JHEP03(2011)015}},
\href{http://www.arXiv.org/abs/1011.2268}{\texttt{arXiv:1011.2268}}.
%%CITATION = ARXIV:1011.2268;%%.

\bibitem{CMS-PAS-JME-16-003}
\href {http://cds.cern.ch/record/2256875?ln=en}{{CMS Collaboration}, ``{Jet
  algorithms performance in 13 TeV data}'',} CMS Physics Analysis Summary
  CMS-PAS-JME-16-003, CERN, 2017.

\bibitem{Sirunyan:2017ezt}
\hrefCMSnoop {}{{CMS Collaboration}, ``{Identification of heavy-flavour jets
  with the CMS detector in pp collisions at 13 TeV}'',} \textit{ JINST}
  \textbf{ 13} (2018) P05011,
  \href{http://dx.doi.org/10.1088/1748-0221/13/05/P05011}{\doi{10.1088/1748-0221/13/05/P05011}},
\href{http://www.arXiv.org/abs/1712.07158}{\texttt{arXiv:1712.07158}}.
%%CITATION = ARXIV:1712.07158;%%.

\bibitem{CMS-DP-2016-038}
\href {https://cds.cern.ch/record/2202971}{{CMS Collaboration}, ``{Tau
  identification in boosted topologies}'',} {CMS Detector Performance Summary}
  CMS-DP-2016-038, CERN, 2016.

\bibitem{Wobisch:1998wt}
\href
  {https://inspirehep.net/record/484872/files/arXiv:hep-ph_9907280.pdf}{M.~Wobisch
  and T.~Wengler, ``{Hadronization corrections to jet cross-sections in deep
  inelastic scattering}'',} in \textit{ {Monte Carlo generators for HERA
  physics. Proceedings, Workshop, Hamburg, Germany, 1998-1999}}, p.~270.
\newblock 1998.
\newblock
\href{http://www.arXiv.org/abs/hep-ph/9907280}{\texttt{arXiv:hep-ph/9907280}}.
\newblock
%%CITATION = HEP-PH/9907280;%%.

\bibitem{Chatrchyan:2012zz}
\hrefCMSnoop {}{{CMS Collaboration}, ``{Performance of tau-lepton
  reconstruction and identification in CMS}'',} \textit{ JINST} \textbf{ 7}
  (2012) P01001,
  \href{http://dx.doi.org/10.1088/1748-0221/7/01/P01001}{\doi{10.1088/1748-0221/7/01/P01001}},
\href{http://www.arXiv.org/abs/1109.6034}{\texttt{arXiv:1109.6034}}.
%%CITATION = ARXIV:1109.6034;%%.

\bibitem{BDT}
T.~Hastie, R.~Tibshirani, and J.~Friedman, ``The elements of statistical
  learning : data mining, inference, and prediction''.
\newblock Springer Series in Statistics. Springer, 2001.

\bibitem{Khachatryan:2015hwa}
\hrefCMSnoop {}{{CMS Collaboration}, ``{Performance of electron reconstruction
  and selection with the CMS detector in proton-proton collisions at $\sqrt{s}
  = 8$\TeV}'',} \textit{ JINST} \textbf{ 10} (2015) P06005,
  \href{http://dx.doi.org/10.1088/1748-0221/10/06/P06005}{\doi{10.1088/1748-0221/10/06/P06005}},
\href{http://www.arXiv.org/abs/1502.02701}{\texttt{arXiv:1502.02701}}.
%%CITATION = ARXIV:1502.02701;%%.

\bibitem{Chatrchyan:2012xi}
\hrefCMSnoop {}{{CMS Collaboration}, ``{Performance of CMS muon reconstruction
  in ${\rm pp}$ collision events at $\sqrt{s} = 7$\TeV}'',} \textit{ JINST}
  \textbf{ 7} (2012) P10002,
  \href{http://dx.doi.org/10.1088/1748-0221/7/10/P10002}{\doi{10.1088/1748-0221/7/10/P10002}},
\href{http://www.arXiv.org/abs/1206.4071}{\texttt{arXiv:1206.4071}}.
%%CITATION = ARXIV:1206.4071;%%.

\bibitem{SVFIT}
\hrefCMSnoop {}{{CMS Collaboration}, ``{Search for neutral MSSM Higgs bosons
  decaying to tau pairs in pp collisions at $\sqrt{s}=7\TeV$}'',} \textit{
  Phys. Rev. Lett.} \textbf{ 106} (2011) 231801,
  \href{http://dx.doi.org/10.1103/PhysRevLett.106.231801}{\doi{10.1103/PhysRevLett.106.231801}},
\href{http://www.arXiv.org/abs/1104.1619}{\texttt{arXiv:1104.1619}}.
%%CITATION = ARXIV:1104.1619;%%.

\bibitem{1742-6596-513-2-022035}
\hrefCMSnoop {}{L.~Bianchini, J.~Conway, E.~K. Friis, and C.~Veelken,
  ``{Reconstruction of the Higgs mass in H$\rightarrow\tau\tau$ events by
  dynamical likelihood techniques}'',} in \textit{ 20th Int. Conf. on Computing
  in High Energy and Nuclear Physics (CHEP2013)}.
\newblock Institute of Physics, Amsterdam, the Netherlands, 2014.
\newblock Journal of Physics: Conference Series 513 (2014) 022035.
  \href{http://dx.doi.org/10.1088/1742-6596/513/2/022035}{\doi{10.1088/1742-6596/513/2/022035}}.

\bibitem{Bianchini:2016yrt}
L.~Bianchini\hrefCMSnoop {}{ {et~al.}, ``{Reconstruction of the Higgs mass in
  events with Higgs bosons decaying into a pair of $\tau$ leptons using matrix
  element techniques}'',} \textit{ Nucl. Instrum. Meth. A} \textbf{ 862} (2017)
  54,
  \href{http://dx.doi.org/10.1016/j.nima.2017.05.001}{\doi{10.1016/j.nima.2017.05.001}},
\href{http://www.arXiv.org/abs/1603.05910}{\texttt{arXiv:1603.05910}}.
%%CITATION = ARXIV:1603.05910;%%.

\bibitem{Garwood}
\hrefCMSnoop {}{F.~Garwood, ``Fiducial limits for the poisson distribution'',}
  \textit{ Biometrika} \textbf{ 28} (1936) 437,
  \href{http://dx.doi.org/10.1093/biomet/28.3-4.437}{\doi{10.1093/biomet/28.3-4.437}}.

\bibitem{Bahr:2008pv}
\hrefCMSnoop {}{M.~Bahr {et~al.}, ``{Herwig++ physics and manual}'',} \textit{
  Eur. Phys. J. C} \textbf{ 58} (2008) 639,
  \href{http://dx.doi.org/10.1140/epjc/s10052-008-0798-9}{\doi{10.1140/epjc/s10052-008-0798-9}},
\href{http://www.arXiv.org/abs/0803.0883}{\texttt{arXiv:0803.0883}}.
%%CITATION = ARXIV:0803.0883;%%.

\bibitem{CMS:lumi}
\href {http://cds.cern.ch/record/2257069?ln=en}{{CMS Collaboration}, ``{CMS}
  luminosity measurement for the 2016 data taking period'',} CMS Physics
  Analysis Summary CMS-PAS-LUM-17-001, CERN, 2017.

\bibitem{Read:2002hq}
\hrefCMSnoop {}{A.~L. Read, ``Presentation of search results: the {$CL_s$}
  technique'',} \textit{ J. Phys. G} \textbf{ 28} (2002) 2693,
\href{http://dx.doi.org/10.1088/0954-3899/28/10/313}{\doi{10.1088/0954-3899/28/10/313}}.
%%CITATION = JPHGB,G28,2693;%%.

\bibitem{AsymptCLs}
\hrefCMSnoop {}{G.~Cowan, K.~Cranmer, E.~Gross, and O.~Vitells, ``Asymptotic
  formulae for likelihood-based tests of new physics'',} \textit{ Eur. Phys. J.
  C} \textbf{ 71} (2011) 1554,
  \href{http://dx.doi.org/10.1140/epjc/s10052-011-1554-0}{\doi{10.1140/epjc/s10052-011-1554-0}},
  \href{http://www.arXiv.org/abs/1007.1727}{\texttt{arXiv:1007.1727}}.
[Erratum: \DOI{10.1140/epjc/s10052-013-2501-z}].
%%CITATION = ARXIV:1007.1727;%%.

\bibitem{Junk:1999kv}
\hrefCMSnoop {}{T.~Junk, ``{Confidence level computation for combining searches
  with small statistics}'',} \textit{ Nucl. Instrum. Meth. A} \textbf{ 434}
  (1999) 435,
  \href{http://dx.doi.org/10.1016/S0168-9002(99)00498-2}{\doi{10.1016/S0168-9002(99)00498-2}},
\href{http://www.arXiv.org/abs/hep-ex/9902006}{\texttt{arXiv:hep-ex/9902006}}.
%%CITATION = HEP-EX/9902006;%%.

\bibitem{CMS-NOTE-2011-005}
\href {http://cds.cern.ch/record/1379837}{{ATLAS and CMS Collaborations, LHC
  Higgs Combination Group}, ``{Procedure for the LHC Higgs boson search
  combination in Summer 2011}'',} Technical Report ATL-PHYS-PUB-2011-11,
  CMS-NOTE-2011-005, CERN, 2011.

\end{thebibliography}\endgroup
\cleardoublepage \appendix\section{The CMS Collaboration \label{app:collab}}\begin{sloppypar}\hyphenpenalty=5000\widowpenalty=500\clubpenalty=5000\vskip\cmsinstskip
\textbf{Yerevan Physics Institute, Yerevan, Armenia}\\*[0pt]
A.M.~Sirunyan, A.~Tumasyan
\vskip\cmsinstskip
\textbf{Institut f\"{u}r Hochenergiephysik, Wien, Austria}\\*[0pt]
W.~Adam, F.~Ambrogi, E.~Asilar, T.~Bergauer, J.~Brandstetter, E.~Brondolin, M.~Dragicevic, J.~Er\"{o}, A.~Escalante~Del~Valle, M.~Flechl, M.~Friedl, R.~Fr\"{u}hwirth\cmsAuthorMark{1}, V.M.~Ghete, J.~Grossmann, J.~Hrubec, M.~Jeitler\cmsAuthorMark{1}, A.~K\"{o}nig, N.~Krammer, I.~Kr\"{a}tschmer, D.~Liko, T.~Madlener, I.~Mikulec, E.~Pree, N.~Rad, H.~Rohringer, J.~Schieck\cmsAuthorMark{1}, R.~Sch\"{o}fbeck, M.~Spanring, D.~Spitzbart, A.~Taurok, W.~Waltenberger, J.~Wittmann, C.-E.~Wulz\cmsAuthorMark{1}, M.~Zarucki
% ...
\vskip\cmsinstskip
% ...
1:  Also at Vienna University of Technology, Vienna, Austria\\
% ...
\end{sloppypar}
\end{document}